\documentclass[12pt,a4paper,notitlepage]{article}
\usepackage[utf8]{inputenc}
\usepackage[english]{babel}
\usepackage[T1]{fontenc}
\usepackage{hyperref}
\usepackage{amsmath}
\usepackage{stmaryrd}
\usepackage{amsfonts}
\usepackage{amssymb}
\usepackage{color} 
\usepackage{caption}
\usepackage{amsmath}
\usepackage{relsize,exscale}
\usepackage{array,multirow,makecell}

\usepackage{bm}
\usepackage{bbold}
\usepackage[toc,page]{appendix}

\setcellgapes{1pt}
\makegapedcells
\newcolumntype{R}[1]{>{\raggedleft\arraybackslash }b{#1}}
\newcolumntype{L}[1]{>{\raggedright\arraybackslash }b{#1}}
\newcolumntype{C}[1]{>{\centering\arraybackslash }b{#1}}
\newcommand{\Tr}{\mathrm{Tr}}

\newtheorem{theorem}{Theorem}
\newtheorem{definition}{Definition}

\newtheorem{remark}{Remark}

\usepackage{enumerate}
\usepackage{subfig}
\usepackage{graphicx}

\newcommand{\beq}{\begin{equation}}
\newcommand{\eeq}{\end{equation}}
\newcommand{\bea}{\begin{eqnarray}}
\newcommand{\eea}{\end{eqnarray}}
\definecolor{mygray}{gray}{0.3}

\newcommand{\bes}{\begin{eqnarray}}
\newcommand{\ees}{\end{eqnarray}}

\newcommand\restr[2]{{
  \left.\kern-\nulldelimiterspace 
  #1 
  \vphantom{\big|} 
  \right|_{#2} 
  }}

\newcommand{\U}{\mathrm{U}}

\usepackage{multicol}
\usepackage{amssymb}

\makeatletter
\begin{document}
\begin{center}
\textbf{\Large{Stochastic melonic kinetics with random initial conditions}}\\

\vspace{15pt}

{\large Bio Wahabou Kpera$^{a}$\footnote{wahaboukpera@gmail.com}, Vincent Lahoche$^b$\footnote{vincent.lahoche@cea.fr}  \,\,and 
Dine Ousmane Samary$^{a,b}$\footnote{dine.ousmanesamary@cipma.uac.bj}
 }
\vspace{15pt}

a)\, International Chair in Mathematical Physics and Applications (ICMPA-UNESCO Chair), University of Abomey-Calavi,
072B.P.50, Cotonou, Republic of Benin

b)\,  Université Paris Saclay, CEA List, Gif-sur-Yvette, F-91191, France

\vspace{0.5cm}
\end{center}
\begin{center}
\textbf{Abstract}
\end{center}
The probability laws associated with random tensors or tensor field theories are traditionally equilibrium distributions. In this paper, we consider a stochastic point of view, and approach the quantization by a Langevin type equation. We especially address the low-temperature behavior of the phase ordering kinetics of a stochastic complex tensor field $T_{i_1\cdots i_d}(x,t)$ of size $N$ and rank $d$ in dimension $D$. The method  we propose use the self averaging property of the tensorial invariants in the large $N$ limit. In this regime, the dynamics is governed by the melonic sector, whose behavior is studied in the quenched limit, where the contractions involving $d-1$ indices self-average around a diagonal matrix proportional to the identity. The following work especially focuses on the cyclic (i.e. non-branching) melonic sector, and we study the way that the system returns to the equilibrium regime regarding the temperature and the shape of the potential. In particular, we provide a general formula for the transition temperature between these regimes. The manuscript is accompanied by numerical simulations to support the theoretical analysis, and essentially aims to open towards this new field of investigation.
\newpage
\tableofcontents
\pagebreak
\bigskip
\section{Introduction}

Phase ordering kinetics is a phenomenon classically describing the growth of an ordered phase as a domain
coarsening for a quenched system, from the homogeneous phase toward a broken phase \cite{Livi}. It has been investigated for $O(N)$ field theory models \cite{Bray,Emmot} from the methods used to solve spin glass dynamics, and physics exhibits exciting relation with the soft $p=2$ spin dynamics \cite{Leticia1, Cugliandolo2, Fyodorov, Dominicsbook, LahocheSamary,Lahoche:2022wal}. In this paper, we investigate the growth of the leading order phase for a stochastic random tensor model (RTM), which in contrast with ordinary phase ordering kinetics does not break the underlying $\U(N)^{\times d}$ symmetry.
\medskip

RTMs were introduced to generalize in higher dimensions the success of random matrix models (RMM) for 2D quantum Einstein gravity \cite{Francesco94}. In 2009, it is shown that the colored RTM admits a tractable $1/N$ expansion involving a generalized degree $\varpi \geq 0$ such that the Feynman amplitudes $A(\mathcal{G})$ for the graph $\mathcal{G} $ scales as $A(\mathcal{G})\sim N^{d-\frac{2}{(d-1)!}\varpi(\mathcal{G})}$ \cite{guruau2017random}. The degree $\varpi$ can be computed from the set of vertices, edges, and faces building the graph $\mathcal{G}$, which is a $2$-simplex rather than an ordinary Feynman graph. The leading order graphs called melons are defined by the condition $\varpi(\mathcal{G})=0$, and follow a recursive definition reminiscent of branched polymers occurring for large $N$ random vectors \cite{zinnvector}. Furthermore, their critical behavior has been investigated as well analytically \cite{Bonzomcritical,Valentindouble}, and confirms the branched polymer phase transition from the critical exponent.  Recently the so called Sachdev–Ye–Kitaev (SYK) quantum mechanics model which consists of N-Majorana fermions with random interactions are showed to admit this same large N-limit \cite{Gurau:2016lzk}-\cite{Maldacena:2016hyu}. Many other works in this direction have also been successful, in particularly: the $O(N)$-tensor model to derive  the fate of the wormholes in a model without quenched disorder with gauge symmetry whose correlation function and thermodynamics in the large
N limit are the same as that of the SYK model \cite{Choudhury:2021nal}-\cite{Choudhury:2017tax}, see also \cite{Bonzom:2015axa}-\cite{Klebanov:2019jup} for similar works. Generally, what makes the RTM power countable is their global $\U(N)$ invariance, and the statistics of RTM follow an exponential law $\rho \sim e^{-S}$, where the classical action $S$ is a sum of $\U(N)^{\times d}$ invariants.
\medskip

In this paper, we give an introduction to the new field of investigation i.e. the stochastic complex tensors $T_{i_1\cdots i_d}(x,t)$ on $\mathbb{R}^{D+1}$, whose dynamics follow a diffusion Langevin type equation, by focusing on the melonic cyclic (i.e. non-branched) sector.
Such a stochastic model has been considered recently for a tensorial group field theory (TGFT) \cite{Lahochestochastic} where a gravitational field was expected to be out of equilibrium with respect to a weakly coupled scalar field playing the role of a physical ''clock '' and materialized by the time variable \cite{orititime}. Our purpose in this manuscript is only to focus on the large time dynamics, where the random tensor is ''quenched" around a mean value. In that limit, the closed equations can be solved with elementary analytic methods, and a transition temperature between a power law versus. an exponential relaxation can be computed. However, these conclusions depend on the shape of the confining potential, and in particular on the order of the roots of the polynomial, which we also highlight.  This kind of model is studied in the framework of vectors field theory in the large $N$-limit  \cite{Bray,Lahoche:2022lmf,Leticia1}. The generalization to the tensor case with a more complicated invariance for a multilinear object is therefore necessary. 
\medskip

\textbf{Outline.} The paper is organized as follows. In section \ref{sec1} we define the model and conventions. In section \ref{sec2} we consider the low-temperature regime for $D=0$ in the quenched limit, and in section \ref{sec3} we extend the method for $D>0$. In both cases, we investigate the large time closed equation expected from the self-averaging assumption and the transition regime toward this limit. In section \ref{sec5} the case of a disordered tensor for $D=0$, the disorder coupling is materialized by the tensor product of Wigner matrices. We conclude in section \ref{sec7}.

\section{Model and conventions}\label{sec1}
We consider a time-dependent complex tensor $\textbf{T}(x,t)$ with rank $d>2$, $t\in \mathbb{R}$ and $x\in \mathbb{R}^D$. We denote by $T_{i_1\cdots i_d}(x,t)\in \mathbb{C}$ the components of the tensor, and the indices $i_1, \cdots, i_d$ run from $1$ to $N$. We furthermore denote by $\bar{\textbf{T}}(x,t)$ the complex-conjugate tensor of $\textbf{T}(x,t)$, and by $\bar{T}_{i_1 \cdots i_d}(x,t)$ its components. The dynamics of the tensor are assumed to follow a Langevin-like equation:
\begin{equation}
\dot{T}_{i_1\cdots i_d}:=-\, \frac{\delta \mathcal{H}}{\delta \bar{T}_{i_1\cdots i_d}}+\partial_x^2 \,T_{i_1\cdots i_d}+\eta_{i_1\cdots i_d}\,.\label{eqLangevin}
\end{equation}
In this equation $\dot{T}_{i_1\cdots i_d}$ is the derivative with respect to $t$, $\partial_x^2$ denotes the standard Laplacian over $\mathbb{R}^D$, $\eta_{i_1\cdots i_d}(x,t)$ is a Gaussian tensor field with correlations defined as
\begin{equation}
\langle \eta_{i_1\cdots i_d}(x,t) \bar{\eta}_{j_1\cdots j_d}(x^\prime,t^\prime) \rangle = \prod_{c=1}^d \frac{\delta_{i_cj_c}e^{-\frac{\vert x-x^\prime\vert^2}{2\Lambda^{-2}}}}{\Lambda^{-D} (2\pi)^{D/2}}F(t-t^\prime)\,,\label{distributioneta}
\end{equation}
where $F(t-t^\prime)=F(t^\prime-t)$ is the memory function, and in the \textit{white noise limit}:
\begin{equation}
F(t-t^\prime)=:T\delta(t-t^\prime)\,,
\end{equation}
the friction coefficient $T$ being interpreted as the temperature for the equilibrium distribution, see below. The UV cut-off $\Lambda$ provides a UV regularization of the theory, which will be clear in the next sections. Note that,
\begin{equation}
\lim_{\Lambda\to\infty}\frac{e^{-\frac{\vert x-x^\prime\vert^2}{2\Lambda^{-2}}}}{\Lambda^{-D} (2\pi)^{D/2}} \to \delta(x-x^\prime)\,.
\end{equation}
The Hamiltonian $\mathcal{H}\equiv \mathcal{H}[\textbf{T},\bar{\textbf{T}}]$ is assumed to be a sum of tensorial invariant, namely:
\begin{align}
\mathcal{H}[\textbf{T},\bar{\textbf{T}}]=\mu\,\vcenter{\hbox{\includegraphics[scale=1]{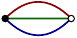}}}+g_{21}\vcenter{\hbox{\includegraphics[scale=1]{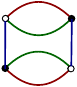}}}+\cdots
+g_{31}\vcenter{\hbox{\includegraphics[scale=1]{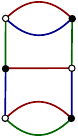}}}+\cdots +g_{41}\vcenter{\hbox{\includegraphics[scale=1]{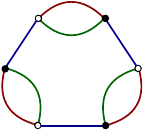}}}+\cdots\,,
\end{align}
where we use standard convention in the RTM literature: black and white vertices materialize respectively fields $\textbf{T}$ and $\bar{\textbf{T}}$, and solid edges materialize Kronecker Dirac delta between indices of different tensors. Furthermore, all the tensor fields are taken at the same point $x$ and at the same time $t$ for each interaction and the Hamiltonian involves a global integral over $\mathbb{R}^D$. \\
We call \textit{a bubble} such a connected graph and note that we set $d=3$ for graphical representations of bubbles in this paper. Note that high valence interactions avoid arbitrary large configurations for the stochastic tensor $\bm T$.
For a fixed initial condition for $t=t_0$, one can consider the probability $P[\textbf{T},\bar{\textbf{T}},t]$ that tensor take the values $(\textbf{T}(t)=\textbf{T},\bar{\textbf{T}}(t)=\bar{\textbf{T}})$ for time $t>t_0$. This probability satisfies a Fokker-Planck equation \cite{ZinnBook}, and for a delta-correlated noise in space dimension zero, the equilibrium probability $\rho[\textbf{T},\bar{\textbf{T}}]:= \lim\limits{t\to \infty} P[\textbf{T},\bar{\textbf{T}},t]$ behaves as,
\begin{equation}
\rho[\textbf{T},\bar{\textbf{T}}] \propto \exp \left(-2\frac{\mathcal{H}[\textbf{T},\bar{\textbf{T}}]}{T}\right)\,,\label{eqdistribution2}
\end{equation}
which corresponds to a standard RTM \textit{at temperature $T$} if the $N$--scaling of couplings constants $\mu$ and $g_{ij}$ labeling each bubble are such that Feynman amplitudes $A(\mathcal{G})$ scale as $N^{d-\frac{2}{(d-1)!}\varpi(\mathcal{G})}$. This can be achieved if coupling constants $g_{b}$ for bubble $b$ scales as $g_{ij}=N^{a(b)}\bar{g}_{ij}$ with:
\begin{equation}
a(b):=-(p-1)(d-1)-\frac{2}{(d-2)!}\varpi(b)\,,
\end{equation}
where the valence $p:=n(b)$ is the number of black vertices in $b$. Note that for melonic bubbles $\varpi(b)\equiv 0$.
\medskip

In the next sections we investigate the large-time behavior in the $N\to \infty$ regime. In particular, we compute the transition temperature investigating, in the quenched regime, the effective dynamics for the self-averaging quantity:
\begin{equation}
r(x,t)\equiv \sum_{i_1\cdots i_d} \frac{\bar{T}_{i_1\cdots i_d}(x,t){T}_{i_1\cdots i_d}(x,t)}{N^d}\,.
\end{equation}
For our purpose in this paper, we focus on the cyclic sector of the melonic family, and we recall some definitions for self-consistency.
\begin{figure}
\begin{center}
$\vcenter{\hbox{\includegraphics[scale=1.2]{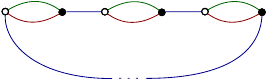}}}$
\end{center}
\caption{Structure of the non-branching melons for $d=3$} \label{fig2}
\end{figure}
\begin{definition}\label{defmelons}
Any melonic bubble $b_p$ of valence $p$ may be deduced from the elementary melon $b_1$:
\begin{equation}
b_1:=\vcenter{\hbox{\includegraphics[scale=1]{graph1.pdf} }}\,,
\end{equation}
replacing successively $(\kappa-1)$-colored edges (including maybe color ‘‘0") by $(d-1)$-dipole, the $(d-1)$-dipole insertion operator $\mathfrak{R}_{c}$ being defined as:
\begin{equation}
\vcenter{\hbox{\includegraphics[scale=1]{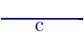} }}\underset{\mathfrak{R}_{c}}{\longrightarrow}\vcenter{\hbox{\includegraphics[scale=1]{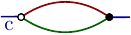} }}\,,
\end{equation}
or formula: $b_p := \left(\prod_{\alpha=1}^{p-1}\mathfrak{R}_{c_\alpha}\right) b_1$. This recursive definition solves the constraint $\varpi(b)\equiv 0$.
\end{definition}
The cyclic melons are then defined as follows:
\begin{definition}\label{defnonbranch}
A non-branching melonic bubble of valence $p$, $b_p^{(\ell)}$ is labeled with a single color index $c\in\llbracket 1,d\rrbracket$, and defined such that:
\begin{equation}
b_p^{(c)}:= \left(\mathfrak{R}_{c}\right)^{p-1}\,b_1\,.
\end{equation}
\end{definition}
Figure \ref{fig2} provides the generic structure of melonic non-branching bubbles in rank $d=3$. We then assume that the Hamiltonian $\mathcal{H}$ reads as
\begin{equation}
\mathcal{H}[\bm T,\bar{\bm T}]=\mu\,\vcenter{\hbox{\includegraphics[scale=1]{graph1.pdf}}}+\sum_{c=1}^d\, \mathcal{V}_c[\bm T,\bar{\bm T}]\,,
\end{equation}
where $\mathcal{V}_c[\bm T,\bar{\bm T}]$ expands as:
\begin{align}
\nonumber\mathcal{V}_c[\bm T,\bar{\bm T}]&=N^{d-1}\,\sum_{p\geq 2}\, \frac{\bar{g}_p}{p N^{p(d-1)}}\,\overbrace{\vcenter{\hbox{\includegraphics[scale=0.8]{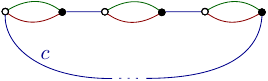}}}}^{p}\\
&=:N^{d-1}\, \sum_{p\geq 1}\, \frac{\bar{g}_p}{p N^{p(d-1)}}\, \Tr (\Phi_c)^p\,,
\end{align}
where entries of the unitary matrices $\Phi_c$ are defined as:
\begin{equation}
(\Phi_c)_{ij}:=\sum_{I,J} \bar{T}_{I} T_J \left(\prod_{\ell \neq c} \delta_{i_\ell j_\ell}\right)\delta_{i_c i}\delta_{j_c j}\,,
\end{equation}
such that $N^d r\equiv \Tr\, (\Phi_c)$.

\section{The $D=0$ model in the quenched limit for a white noise}\label{sec2}

This section is devoted to the dynamic aspects of the zero-dimensional model. Equation \eqref{eqLangevin} can be investigated for a large time, and with some approximations that we will detail, the radius $r(t)$ satisfies a closed equation that can be studied analytically \cite{Bray}. The way that the system converges toward equilibrium for late time can be investigated as well. The zero-dimensional model has no particular interest in itself, and the aim of this section is essential to introduce the general strategy that we will implement in the next two sections.

\subsection{Static limit}
The equilibrium points of our model are given by the relation:
\begin{align}
 0= \frac{\partial \mathcal{H}}{\partial \bar{T}_{i_1\cdots i_d}} = \mu\, \vcenter{\hbox{\includegraphics[scale=1]{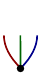}}}+\sum_{c=1}^d\, \Bigg( \,g_2\,\vcenter{\hbox{\includegraphics[scale=1]{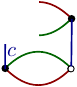}}}
+\,g_3\,\vcenter{\hbox{\includegraphics[scale=0.8]{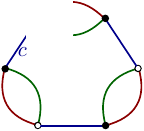}}}+\cdots\Bigg)\,,\label{equationsaddle}
\end{align}
where the deleted nodes correspond to the partial derivatives with respect to the field $\Phi_c$. The vacuum solution can be chosen to be $\U(N)$ invariant, namely\footnote{A corollary of the standard Schur Lemma.}:
\begin{equation}
(\Phi_c)_{ij}(t)=\kappa_c(t) \,\delta_{ij}\,,\label{diagassumption}
\end{equation}
which can be easily justified in the large $N$ limit due to our random initial condition \eqref{initialcondition}. We furthermore assume that color permutation symmetry is unbroken, and $\kappa_c\equiv \kappa,\,\,\forall\, c$. In that way, equation \eqref{equationsaddle} becomes:
\begin{equation}
0\equiv T_I \times \frac{\partial}{\partial \kappa}\, \left( \mu \kappa + d\,\sum_{p\geq 2}\, \frac{g_p}{p}\, \kappa^p\right)=:T_I\times \frac{\partial U(\kappa)}{\partial \kappa}\,,
\end{equation}
where $I:={i_1\cdots i_d}$. Hence, assuming $T_I \neq 0$, the equilibrium points are defined by the minima of the potential $U(\kappa)$. For the quartic model, this point is given by:
\begin{equation}
\kappa=-N^{d-1}\left(\frac{\mu}{\bar{g}_2 d}\right)=\mathcal{O}(N^{d-1})\,.
\end{equation}
If we assume that the system does not break the $\U(N)$ symmetry dynamically and for large $N$ $\Phi_c$ \textit{self average} around a diagonal colored symmetric matrix during the time evolution (see remark \ref{remark1} at the end of this section), the Langevin equation \eqref{eqLangevin} can be approached as:
\begin{equation}
\dot{T}_{i_1\cdots i_d}=-\, {T}_{i_1\cdots i_d}\, U^\prime(\kappa)+\eta_{i_1\cdots i_d}\,,\label{eqLangevin2}
\end{equation}
where $\kappa$ is assumed to be replaced by its averaged value, such that each component of $\bm T$ decouples from the others. The previous equation can be formally solved as:
\begin{align}
{T}_{i_1\cdots i_d}(t)={T}_{i_1\cdots i_d}(0)\rho(t)+\int_0^t dt^\prime\, \eta_{i_1\cdots i_d}(t^\prime)\, \frac{\rho(t)}{\rho(t^\prime)}\,,\label{formalsol}
\end{align}
with:
\begin{equation}
\rho(t):=e^{-g(t)}\,,\qquad g(t):=\int_0^t dt^\prime\, U^\prime(\kappa(t^\prime))\,.
\end{equation}
In equation \eqref{formalsol}, the time $t=0$ has to be understood not as a true initial condition but as an arbitrary time, far from the true origin of time. Because we have no information about the true behavior of the tensor at $t=0$ for a given draw, we impose it to be randomly distributed accordingly with the centred Gaussian distribution\footnote{We define arbitrary $D$ here, despite this section focusing on the case $D=0$.}:
\begin{equation}
\langle \bar{T}_{i_1\cdots i_d}(x,0) T_{j_1\cdots j_d}(x^\prime,0) \rangle_0 = \Delta\, \prod_{c=1}^d \delta_{i_cj_c} \frac{e^{-\frac{\vert x-x^\prime\vert^2}{2\Lambda^{-2}}}}{\Lambda^{-D} (2\pi)^{D/2}}\,,\label{initialcondition}
\end{equation}
where $\vert x \vert$ is the standard Euclidean norm, $\langle X \rangle_0$ means averaging over initial conditions and $\Lambda$ is the same UV cutoff as for the noise $\bm \eta(x,t)$. Hence we get the closed equation:
\begin{equation}
\boxed{r(t) G(t)= \Delta\,+\, T \int_0^tdt^\prime\, G(t^\prime)\,,}
\end{equation}
\medskip
where $G(t):=e^{2g(t)}$. For large time, we expect that $r(t)$ reaches one of the equilibrium points of $U(\kappa)$, i.e., that $U'(\kappa(t))\to 0$ for $t$ large enough (see Section \ref{sec3}). From this expectation, the previous equation looks like a closed integral equation for $G(t)$, which can be easily solved asymptotically.
\begin{itemize}
\item \textit{For $T=0$}, we assume $r(t) \to r_0 \neq 0$ for $t$ large enough, such that $U'(N^{d-1} r_0)=0$. Hence, $G(t)\to \Delta/r_0$, and $U'(\kappa(t)) \to 0$, in agreement with our assumption. For $r_0=0$, if $\Delta \neq 0$, then $G(t)$ has to diverges with $t$, and $\langle T_{i_1,\cdots,i_d} \rangle \to 0$.

\item \textit{For $T\neq 0$}, differentiating the closed equation concerning $t$, we get $\dot{G}\sim (T/r_0) G(t)\,\to G(t)\sim e^{(T/r_0)t}$, therefore $U'(\kappa(t)) \to (T/2r_0)$, and the system is repelled from the equilibrium point due to thermal fluctuations.
\end{itemize}

\subsection{Equilibrium dynamics}

The time evolution toward the asymptotic regime can be investigated with a little improvement of the previous argument. Indeed, assuming the validity of the quenching regime, and because from the definition:
\begin{equation}
U^\prime(\kappa(t))=\frac{1}{2} \frac{\dot{G}}{G}\,,\label{defUprime}
\end{equation}
we have for the quartic model:
\begin{equation}
\frac{1}{2} \dot{G} = \mu G(t)+ d \bar{g}_2\, \left(\Delta\,+\, T \int_0^tdt^\prime\, G(t^\prime)\right) \,.\label{equationdynamic}
\end{equation}
Deriving the equation with respect to $t$, we have:
\begin{equation}
\ddot{G}(t)-2\mu \dot{G}(t)-2d \bar{g}_2 TG(t)=0\,.
\end{equation}
This is a linear homogeneous differential equation of second order, that can be easily solved as a combination of exponential:
\begin{equation}
G(t)=\alpha e^{i \omega_+ t}+ \beta e^{i \omega_- t}\,.\label{equationG}
\end{equation}
where $\omega_\pm$ is the solution of the characteristic polynomial, namely:
\begin{equation}
\omega^2+2i \mu \omega +2d \bar{g}_2 T=0\,, \label{eqFourier}
\end{equation}
explicitly:
\begin{equation}
\omega_\pm=-i\mu \left(1\pm \sqrt{1+\frac{2d \bar{g}_2 T}{\mu^2}}\right)\,.
\end{equation}
We have two initial conditions. First $G(0)=1$ leading to $\alpha+\beta=1$. Secondly $\dot{G}(0)=2 U^\prime (\kappa(0))$, i.e.
\begin{equation}
\dot{G}(0)=2 (\mu+ d\bar{g}_2 \Delta)=-2\mu \left(\frac{\Delta}{r_0}-1\right)
\end{equation}
(see remark \ref{remark1} below) and where $r_0:=-\mu/(\bar{g}_2 d)$. Hence, we must have:
\begin{equation}
i \omega_+ \alpha+ i \omega_- (1-\alpha)=-2\mu \left(\frac{\Delta}{r_0}-1\right)\,,
\end{equation}
in other words,
\begin{equation}
\alpha=-\frac{2 \left(\frac{\Delta}{r_0}-1\right)+ \left(1- \sqrt{1+\frac{2d \bar{g}_2 T}{\mu^2}}\right)}{2\sqrt{1+\frac{2d \bar{g}_2 T}{\mu^2}}}\,,
\end{equation}
and especially for $T=0$:
\begin{equation}
\alpha=1-\frac{\Delta}{r_0}\,,\qquad \beta=\frac{\Delta}{r_0}\,.
\end{equation}
For $T=0$, we find an exponential law $G(t) = \alpha e^{2\mu t}+\beta$ and the behavior of the system depends on the sign of $\mu$. For $\mu<0$, $G(t)\to \Delta/r_0$ exponentially, as expected. For $\mu>0$, $U^{\prime}(\kappa(t)) \to \mu$ for large time, and then $\kappa(t) \to 0$. Note that because $G(t)$ is positive, we must have $\Delta < r_0$ if we expect that the solution holds for all times. If $\Delta > r_0$, $G(t)$ vanishes for $t=\tau$,
\begin{equation}
\tau= \frac{1}{2\mu}\ln \left( \frac{1}{1-r_0/\Delta}\right)\,.
\end{equation}
For $T\neq 0$, the two solutions $\omega_-$ and $\omega_+$ have different signs, and in particular $i \omega_- < 0$. For small $T$:
\begin{equation}
\omega_-= -i\frac{T}{r_0}+\mathcal{O}(T^2)\,,
\end{equation}
and:
\begin{equation}
G(t) \sim \alpha e^{2\mu t}+\beta e^{(T/r_0)t}\,.
\end{equation}
Hence for $\mu<0$, we recover the asymptotic behavior expected from the previous analysis. As discussed previously for $T\neq 0$, the system fails to reach the equilibrium point of the potential. Nevertheless, it reaches an equilibrium regime that we can characterize. Indeed, from \eqref{equationG}, we have for large time:
\begin{equation}
G(t) \sim \min_{i \omega_\pm} e^{i \omega_{\pm}t}\,,
\end{equation}
or:
\begin{equation}
2 U^{\prime}(\kappa(t)) \sim \min_{i \omega_\pm} \mu \left(1\pm \sqrt{1+\frac{2d \bar{g}_2 T}{\mu^2}}\right)\,.
\end{equation}
To investigate the meaning of this equation, let us focus on the case $\mu > 0$, in which the previous equation can be solved as:
\begin{equation}
\kappa(t)\to \mu\frac{\sqrt{1+\frac{2d \bar{g}_2 T}{\mu^2}}-1}{2d \bar{g}_2}\,. \label{asymptotickappa}
\end{equation}
Now let us consider the equilibrium distribution \label{eqdistribution}. The melonic equilibrium solution can be investigated from the standard methods in RTM literature \cite{guruau2017random} using Schwinger-Dyson equations, which we recall briefly here in this context for self-consistency. The partition function of the equilibrium state reads:
\begin{equation}
Z_{\text{eq}}:= \int d \bm T d\bar{\bm T} \, e^{-2 \mathcal{H}[\bm T \bar{\bm T}]/T}\,,
\end{equation}
where $d \bm T d\bar{\bm T} := \prod_{i_1,\cdots i_d} dT_{i_1\cdots i_d} d\bar{T}_{i_1\cdots i_d}$ is the standard Lebesgue measure, and $\bm{T}$ is now time independent. Let us consider the Schwinger-Dyson equation:
\begin{equation}
0=\sum_I \int d \bm T d\bar{\bm T}\, \frac{\partial}{\partial T_I} \left(\,T_I\, e^{-2 \mathcal{H}[\bm T \bar{\bm T}]/T}\right)\,.
\end{equation}
Computing the derivative, and dividing the resulting equation by $N^d Z_{\text{eq}}$, we arrive to the equation:
\begin{equation}
1-\frac{2\mu}{T} Q- \frac{2d\bar{g}_2}{T} Q^2\asymp 0\,,\label{largeNSDE}
\end{equation}
where $Q$ is defined as:
\begin{equation}
Q:= \frac{1}{N^d}\, \frac{1}{Z_{\text{eq}}} \int d \bm T d\bar{\bm T}\, \left(\sum_I \bar{T}_I T_I \right)\, e^{-2 \mathcal{H}[\bm T \bar{\bm T}]/T}\,,
\end{equation}
and used of the large $N$ self averaging condition \cite{Valentin2011}:
\begin{equation}
\frac{1}{N^d Z_{\text{eq}}} \int d \bm T d\bar{\bm T} \, \mathcal{V}_c[\bm T,\bar{\bm T}] e^{-2 \mathcal{H}[\bm T \bar{\bm T}]/T} \asymp \sum_{p\geq 1} \frac{\bar{g}_p}{p} Q^p\,.
\end{equation}
The solution of \eqref{largeNSDE} is:
\begin{equation}
Q= \mu\frac{\sqrt{1+\frac{2d \bar{g}_2 T}{\mu^2}}-1}{2d \bar{g}_2}\,,\label{eqsol}
\end{equation}
where we selected only the solution that goes toward $T/2\mu$ as $\bar{g}_2$ vanishes. This shows that $\kappa(t) \to Q$ for large $t$, and that the system goes toward the equilibrium regime described by the equilibrium state \eqref{eqdistribution2}. Note that equilibrium solution \eqref{eqsol} is defined below the critical value:
\begin{equation}
g_c=-\frac{\mu^2}{2d T}\,,
\end{equation}
at which the free energy $f_{\text{eq}}:=\ln Z_{\text{eq}}$ becomes non-analytic,
\begin{equation}
f_{\text{eq}} \sim \vert g-g_c\vert^{3/2}=:\vert g-g_c \vert^{2-\theta}
\end{equation}
corresponding to the entropy exponent $\theta=1/2$. Finally, note that the leading order Schwinger Dyson equation means that:
\begin{equation}
2 U^\prime(\kappa)=\frac{T}{\kappa}=\max (i\omega_\pm)\,,
\end{equation}
hence, $G(t)\sim e^{(T/\kappa)t}$ rather than $e^{(T/r_0) t}$.

\subsection{Low temperature regime}\label{lowtempregime}

The result can be generalized for potentials of the form:
\begin{equation}
U^{\prime}(\kappa)=\frac{1}{2}(\kappa-\gamma)R(\kappa)\,,
\end{equation}
where the zeros of the function $R(\kappa)$ are assumed to be different and far enough to the isolated zero $\kappa=\gamma$. If the system is found initially in the vicinity of the isolated zero, and if we assume that the system converges toward some equilibrium value $\kappa_\infty$ for late time, close enough to the value
$\kappa=\gamma$ (see also Section \ref{sec3}, paragraph $b$), the equation for $\dot{G}$ reads approximately:
\begin{equation}
\dot{G}\approx G(t) \left(\frac{\Delta\,+\, T \int_0^tdt^\prime\, G(t^\prime)}{G(t)}-\gamma \right)R(\kappa_\infty) \,,
\end{equation}
which can be solved again as a combination of exponential with:
\begin{equation}
i\omega_{\pm}=R(\kappa_\infty) \frac{\gamma}{2} \left(\pm \sqrt{1+\frac{4T}{\gamma^2R(\kappa_\infty)}}-1 \right)
\end{equation}
The late time behavior corresponds to the $\min i\omega_{\pm}$, and the equation fixing the value of $\kappa_\infty$ is:
\begin{equation}
R(\kappa_\infty) \frac{\gamma}{2} \left(\sqrt{1+\frac{4T}{\gamma^2R(\kappa_\infty)}}-1 \right) = (\kappa_\infty-\gamma)R(\kappa_\infty)\,,
\end{equation}
which can be rewritten as:
\begin{equation}
\kappa_\infty (\kappa_\infty-\gamma) R(\kappa_\infty)=T\,,
\end{equation}
which is nothing but the leading order Schwinger Dyson equation for large $N$.

\medskip

It seems that the ability of the system to reach the equilibrium distribution depends strongly on the shape of the potential. As an example, let us investigate the case of a potential with twice degenerate vacua:
\begin{equation}
\frac{\partial U(\kappa)}{\partial \kappa}=\frac{1}{2} (\kappa-\gamma)^2\,.
\end{equation}
The equation for $\dot{G}$ reads:
\begin{equation}
\dot{G}=G \left(\frac{\Delta\,+\, T \int_0^tdt^\prime\, G(t^\prime)}{G(t)}-\gamma\right)^2
\end{equation}
Because $G$ is positive, the right-hand side is positive at all times, and $G$ can only grow over time. For $T$ small enough, one can expand the right-hand side up to linear order in $T$, and for late time, one can neglect $\Delta/G$ with respect to contributions of order $G^0$ and $G^1$. We have:
\begin{align}
\dot{G}=\Bigg(\gamma^2 G(t)-2\gamma T \int_0^t G(t^\prime) dt^\prime + 2\Delta \left( T\, \frac{\int_0^t G(t^\prime) dt^\prime}{G(t)}-\gamma \right) \Bigg)\,.
\end{align}
Assuming that $G(t)$ expands in power series of $T$:
\begin{equation}
G(t)=\sum_{n=0}^\infty\, T^n G_{n}(t)\,,
\end{equation}
we get for $G_0$:
\begin{equation}
G_0(t)= C_0\, e^{\gamma^2 t} + \frac{2\Delta}{\gamma}\,.
\end{equation}
In the same way, the equation for $G^{(1)}$ reads:
\begin{equation}
\dot{G}_1=\gamma^2 G_1-2\gamma \int_0^t G_0(t^\prime) dt^\prime+2\Delta \frac{\int_0^t G_0(t^\prime) dt^\prime}{G_0(t)}\,,
\end{equation}
Which can be solved as:
\begin{equation}
G_1(t)=\Bigg(C_1- 2\int_0^t dt^\prime \int_0^{t^\prime} dt^{\prime\prime} G_0(t^{\prime\prime}) \Bigg[\gamma - \frac{\Delta}{G_0(t^\prime)} \Bigg]\Bigg) e^{\gamma^2 t}\,.
\end{equation}
The numerical constants $C_1$ and $C_0$ will be fixed by the initial condition $G(0)=1$, leading to:
\begin{equation}
C_0+T\, C_1+ \frac{2\Delta}{\gamma} =1
\end{equation}
and $G(t)$ reads:
\begin{align}
 G(t)=&e^{\gamma^2 t}\Bigg(\left(1-\frac{2\Delta}{\gamma}\right)- 2T\,\int_0^t dt^\prime e^{-\gamma^2 t^\prime}\int_0^{t^\prime} dt^{\prime\prime} G_0(t^{\prime\prime}) \Bigg[\gamma -\frac{\Delta}{G_0(t^\prime)} \Bigg]\Bigg)\,.
\end{align}
For a later time, the right-hand side can be computed exactly, and we have:
\begin{equation}
G(t) \approx e^{\gamma^2 t}\Bigg(\left(1-\frac{2\Delta}{\gamma}\right)-2T \left(\frac{A}{12 \gamma ^4 \Delta }+\frac{1-2\Delta/\gamma}{\gamma}t \right) \Bigg)\,,
\end{equation}
where:
\bea
 A:&=&\Delta \left(\left(\pi ^2-6\right) \gamma -2 \left(\pi ^2-12\right) \Delta \right)+6 \Delta (\gamma -2 \Delta ) \text{Li}_2\left(1-\frac{\gamma }{2 \Delta }\right)
\cr
&+&3 (\gamma -2 \Delta ) \left(\Delta \log ^2\left(\frac{\gamma }{2 \Delta }-1\right)+\gamma \log \left(1-\frac{2 \Delta }{\gamma }\right)\right)\,,
\eea
where $\text{Li}_s(z)$ is the standard poly-logarithm function. Computing the first derivative with respect to $t$, a straightforward calculation leads to\footnote{The computation requires that $t$ remains not too large, $t \lesssim T/\gamma$.}:
\begin{equation}
\frac{\dot{G}}{G} \approx \gamma^2-\frac{2T}{\gamma}+\mathcal{O}(T^2)\,.
\end{equation}
Hence, for $T$ small enough, the asymptotic equation fixing $\kappa$ is:
\begin{equation}
\gamma^2-\frac{2T}{\gamma} \approx (\kappa-\gamma)^2\,,
\end{equation}
which have two solutions:
\begin{equation}
\kappa = \frac{T}{\gamma^2}\,,\quad \textit{or}\quad \kappa= 2\gamma-\frac{T}{\gamma^2}\,.
\end{equation}
In contrast, the leading order Schwinger Dyson equation for the equilibrium theory reads:
\begin{equation}
Q U^\prime (Q)=T\,,
\end{equation}
which, for small $T$ solves as:
\begin{equation}
Q\approx \gamma \pm \sqrt{\frac{T}{\gamma}}\,, \quad \text{or}\quad Q \approx \frac{T}{\gamma^2}\,.\label{zerospert}
\end{equation}
This result shows that, at least for $T$ small enough, equilibrium dynamics selects only one of the three solutions of the equilibrium distribution, the vacuum such that $Q= \mathcal{O}(T)$. One can expect that this concerns only zeros with even degeneracy, but a simple argument shows that it concerns odd potential as well. To show this, let us consider the potential:
\begin{equation}
\dot{G}=G(t) \left(\frac{\Delta\,+\, T \int_0^tdt^\prime\, G(t^\prime)}{G(t)}-\gamma\right)^{n}\,. \label{equationn}
\end{equation}
In contrast with the previous case, the right-hand side is not positive, and one may expect that the equation converges toward a finite value, say $\kappa_\infty$. Hence, assuming the system is closed to this limit:
\begin{equation}
\kappa(t)=\kappa_\infty+\varepsilon(t)\,.
\end{equation}
At first order in $\varepsilon(t)$, we have:
\begin{equation}
\dot{G}=n (\kappa_\infty-\gamma)^{2n} \left(\Delta\,+\, T \int_0^tdt^\prime\, G(t^\prime)-\kappa_\infty G(t)\right)\,,
\end{equation}
corresponding to the frequencies:
\begin{align}
 i\omega_{\pm}= n (\kappa_\infty-\gamma)^{2n} \frac{\kappa_\infty}{2}\left(\pm \sqrt{1+\frac{4T}{\kappa_\infty^2 n (\kappa_\infty-\gamma)^{2n} }}-1 \right)\,.
\end{align}
For late time, the equation for $\kappa_\infty$ can be rewritten as:
\begin{equation}
\kappa_{\infty} i\omega_+\approx T\,,
\end{equation}
which matches with the leading order Schwinger Dyson equation for $T$ small enough. Hence, these simple arguments seem to indicate that potentials having degenerate zeros converge toward equilibrium for late time, i.e. the asymptotic probability distribution agrees with the predictions of the equilibrium distribution \eqref{eqdistribution2} for $T$ small enough. In the next section, we confirm these statements with numerical simulations for the late-time behavior.
\medskip

\subsection{Numerical study}
\begin{figure}
\begin{center}
\includegraphics[scale=0.6]{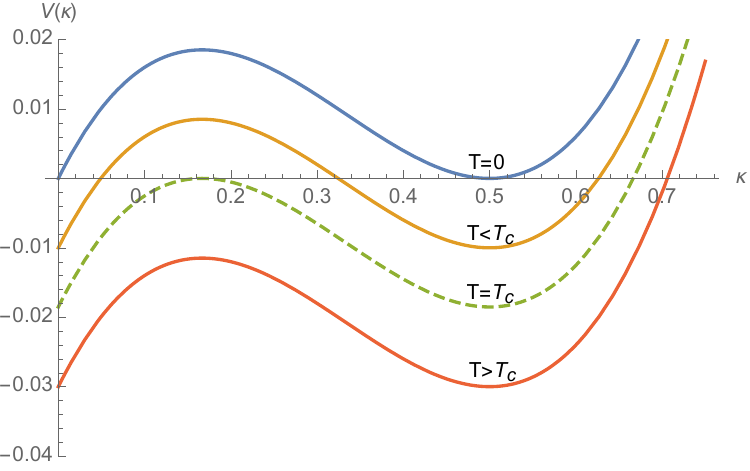}
\includegraphics[scale=0.6]{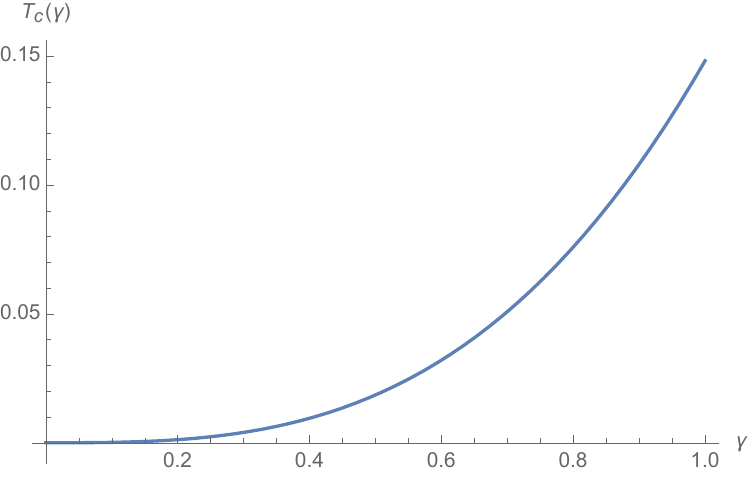}
\end{center}
\caption{Shape of the potential $V(\kappa)$ for some temperature above and below the critical temperature with $\gamma = 0.5$ and critical temperature dependency on $\gamma$.}\label{figPotential}
\end{figure}

The equation for $\dot{G}$ can be investigated numerically. In this section, we focus on the case of purely degenerate vacua, of the form \eqref{equationn}. First, let us consider the quadratic case. If equilibrium is reached, the leading contribution to $G(t)$ for a late time has to behave as \footnote{Let $G(t)=A\, e^{B t}$. Assuming $t$ large enough, $G(t)$ is a solution of the equation of motion if :
\begin{equation}
B= \left(\frac{T}{B}-\gamma\right)^n\,.
\end{equation}
Hence, if $B=T/\kappa_\infty$, the equation reads $\kappa_\infty(\kappa_\infty-\gamma)^n=T$, which is nothing but the leading order Schwinger Dyson equation.}:
\begin{equation}
G(t) \sim \alpha \,\exp \left( \frac{T}{\kappa_\infty} t \right),
\end{equation}
where $\kappa_\infty$ is a zero of the potential $V(\kappa):=\kappa(\kappa-\gamma)^2-T$. There are two regimes, depending on the temperature. For $T<T_c(\gamma)$, there are three real zeros, and for $T\ll 1$ these zeros display as equation \eqref{zerospert} shows, with a zero of order $T$ and two zeros of order $\gamma$. For $T>T_c(\gamma)$ in contrast, the potential has only one real zero, identified from continuity with $T$ as the larger zero among the three real zeros occurring from the low-temperature regime. Figure \ref{figPotential} shows the behavior of the potential $V(\kappa)$ and $T_c(\gamma)$, which is explicitly given by:
\begin{equation}
T_c(\gamma)=\frac{4\gamma ^3}{27}\,.
\end{equation}
\medskip

\begin{figure}
\begin{center}
\includegraphics[scale=0.6]{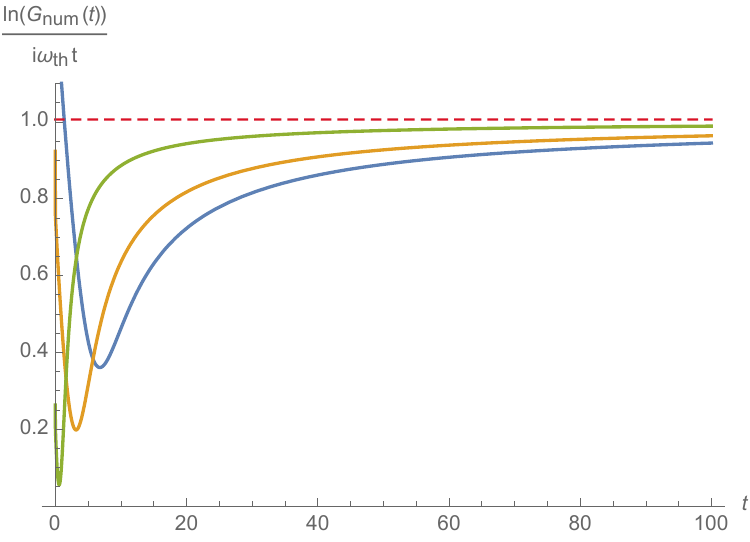}
\includegraphics[scale=0.6]{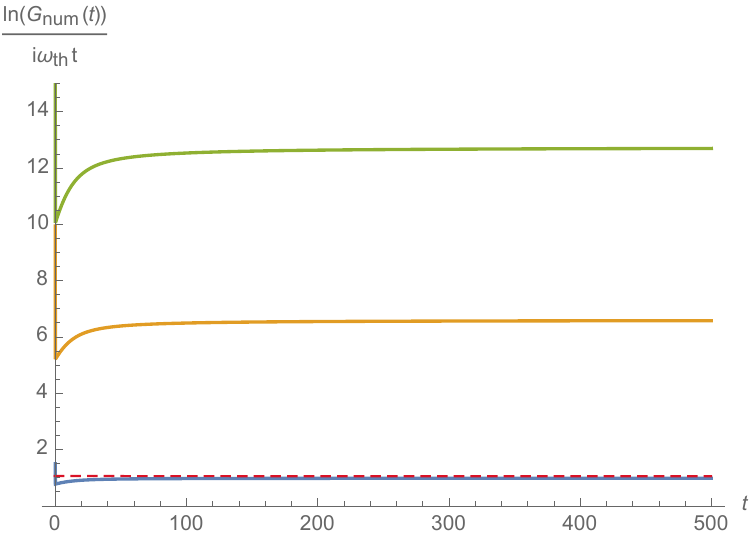}
\end{center}
\caption{Behavior of $\ln G_{\text{num}}/(i \omega_{\text{th}}t)$ respectively for $T>T_c$ (on the left) and for $T<T_c$ (on the right).}\label{figPlot4}
\end{figure}

\begin{figure}
\begin{center}
\includegraphics[scale=0.7]{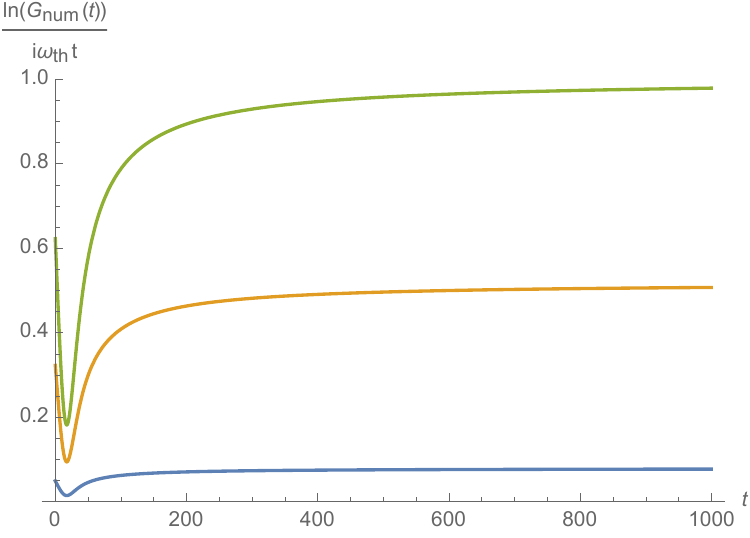}
\end{center}
\caption{Behavior of $\ln G_{\text{num}}/(i \omega_{\text{th}}t)$ below the critical temperature for $\Delta=0.4$ and $\gamma=0.5$.}\label{figPlot5}
\end{figure}

Let us denote as $G_{\text{num}}$ the numerical solution of equation \eqref{equationn} for $n=2$. Figure \ref{figPlot4} shows the behavior of $\ln (G_{\text{num}}(t))/(i \omega_{\text{th}}t)$ above and below the critical temperature $T_c \approx 0.019$ for $\gamma=0.5$ and $\Delta=0.1$, the theoretical value $i \omega_{\text{th}}$ being equals to $T/\kappa_\infty$, such that $V(\kappa_\infty)=0$. As we can show on the Figure \ref{figPlot4} for $T=0.1$, $0.2$ and $1$, $G_{\text{num}}(t)\sim e^{i \omega_{\text{th}} t}$ for $t$ large enough. The same thing occurs below the critical temperature, and the second diagram in Figure \ref{figPlot4} shows the behavior of $\ln (G_{\text{num}}(t))/(i \omega_{\text{th}}t)$ for $T=0.01$ for each of the three solutions of the equation $V(\kappa_\infty)=0$. Among the three solutions, only one converges to $1$ for late time, the closed one to the initial condition $\kappa(t=0)=\Delta$. Figure \ref{figPlot5} illustrate how the system flips between the two stable vacua (i.e. the smallest and the largest), for $\Delta=0.4$. Colors have the same meaning in Figure \ref{figPlot4} and \ref{figPlot5}: the blue curve corresponds to the smallest zero $\kappa_\infty \approx T/\gamma$ and the green curve to the largest zero $\kappa_\infty \approx \gamma + \sqrt{T/\gamma}$. For $\Delta=0.1$ the system reaches the smallest vacuum, and for $\Delta=0.4$ the system reaches the larger vacuum.
\medskip

After a transition regime, the relaxation toward equilibrium can be numerically investigated, and, indeed:
\begin{equation}
\frac{\ln (G_{\text{num}})}{i \omega_{\text{th}}t} \sim 1 - \frac{\alpha}{t}\,.
\end{equation}

\begin{figure}
\begin{center}
\includegraphics[scale=0.7]{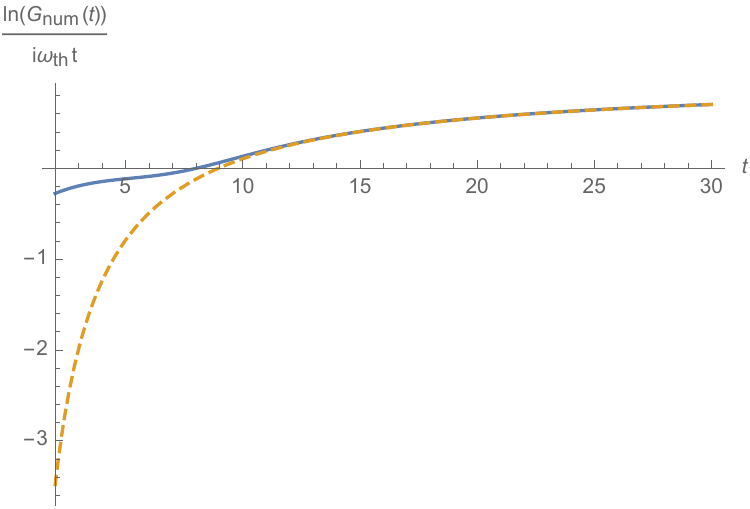}
\end{center}
\caption{Behavior of $\ln (G_{\text{num}}(t))/(i \omega_{\text{th}}t)$ for $n=3$. The dotted edge corresponds to $f(t):=1 - 8.9682/t$.}\label{figPlot6}
\end{figure}

Similar behavior is observed for $n>2$. In general, there are a positive and a negative zero for $V(\kappa)$, and the value toward which $\ln (G_{\text{num}}(t))/(i \omega_{\text{th}}t)$ converges is compatible with Schwinger-Dyson equation for equilibrium state. Figure \ref{figPlot6} shows the behavior of $\ln (G_{\text{num}}(t))/(i \omega_{\text{th}}t)$ for $\Delta=0.1$, $\gamma=0.5$ and $T=0.1$. Numerically, the system reaches the equilibrium state and:
\begin{equation}
G_{\text{num}}(t) \sim 0.396 \, \exp \left( 0.103 t \right)\,.
\end{equation}

\begin{remark}\label{remark1}
The assumption \eqref{diagassumption} can be easily justified along the dynamics from the following argument. Let us compute $\dot{\lambda}_\mu$, the rate of the $\mu$-th eigenvalue of $(\Phi_c)_{ij}$. Because $\Phi_c$ is hermitian, it has $N$ real eigenvalues and admits $N$ normalized and orthogonal eigenvectors $\{u_i^{(\mu)}\}$, $\mu\in \llbracket 1,N \rrbracket$, such that:
\begin{equation}
\sum_{j=1}^N (\Phi_c)_{ij}u_j^{(\mu)}=\lambda_\mu u_i^{(\mu)}\,.
\end{equation}
It is easy to check that:
\begin{align}
\nonumber\sum_{i,j=1}^N\, u_i^{(\mu)}u_j^{(\mu)}(\dot{\Phi}_c)_{ij}&=\dot{\lambda}_\mu-\lambda_\mu \frac{d}{dt}\sum_{i=1}^N u_i^{(\mu)}u_i^{(\mu)}\\
&= \dot{\lambda}_\mu-\lambda_\mu \frac{d}{dt}1\equiv\dot{\lambda}_\mu\,.
\end{align}
The rate $\dot{\lambda}_\mu$ has then the general structure:
\begin{equation}
\dot{\lambda}_\mu=F_\mu\,,
\end{equation}
where $F_\mu$ is the projection of the right-hand side of the Langevin equation. Now, let us investigate the large $N$ distribution $P_0(\vec \Phi_0)$ for the initial distribution of $\Phi_{0,c}:=\Phi_c(t=0)$, where $\vec{\Phi}_0:=\{\Phi_{0,c}\}$. Because \eqref{initialcondition}, we have:
\begin{equation}
\langle (\Phi_{0,c})_{ij} \rangle_0 = N^{d-1}\Delta \delta_{ij}\,.
\end{equation}
Computing the $2$-point correlation, we have:
\begin{align}
\langle (\Phi_{0,c})_{ij} (\Phi_{0,c})_{kl} \rangle_0= \langle (\Phi_{0,c})_{ij} \rangle_0\langle (\Phi_{0,c})_{kl} \rangle_0 +\langle (\Phi_{0,c})_{ij} (\Phi_{0,c})_{kl} \rangle_{0,C}\,,
\end{align}
where the last piece is the connected contribution, which can be easily computed from \eqref{initialcondition}. This leads to:
\begin{align}
\nonumber \langle (\Phi_{0,c})_{ij} (\Phi_{0,c})_{kl} \rangle_{0,C}&=\Delta^2 \sum_{I/c,J/c} \delta_{I/c,J/c} \delta_{il}\delta_{jk}\\
&= N^{d-1}\Delta^2\delta_{il}\delta_{jk}\,,
\end{align}
where $I/c \in \llbracket 1,N \rrbracket^{d-1}$ includes all indices excepts the one of color $c$. The connected contribution is sub-leading (with a factor $1/N^{d-1}$) concerning the disconnected piece, and in the large $N$ limit we have:
\begin{equation}
\langle (\Phi_{0,c})_{ij} (\Phi_{0,c})_{kl} \rangle_0\simeq \langle (\Phi_{0,c})_{ij} \rangle_0\langle (\Phi_{0,c})_{kl} \rangle_0\,.
\end{equation}
In the same way to find:
\begin{equation}
\langle \prod_{\ell=1}^N(\Phi_{0,c})_{i_\ell j_\ell} \rangle_0\simeq \prod_{\ell=1}^N\langle (\Phi_{0,c})_{i_\ell j_\ell} \rangle_0\,,
\end{equation}
and the probability $P_0$, at leading order in $N$ corresponds to a delta-distribution:
\begin{equation}
P_0(\vec \Phi_0)\asymp\prod_{c=1}^d \delta ((\Phi_0)_c-N^{d-1}\Delta \,\mathrm{Id})\,,
\end{equation}
where $\mathrm{Id}$ is the $N\times N$ identity matrix. At the initial time, the distribution is localized around a diagonal matrix with equal entries. Hence, in the rate equation for $\dot{\lambda}_\mu$, initial conditions are the same for all eigenvalues, and they remain identical for all time. The argument easily generalizes for $D>0$.
\end{remark}

\section{White noise limit for $D>0$: UV regularized theory below critical temperature}\label{sec3}

For $D>0$, the Laplacian contribution in equation \eqref{eqLangevin} modify the effective large $N$ dynamics given by equation \eqref{eqLangevin2} as follows:
\begin{equation}
\dot{T}_{i_1\cdots i_d}=-\, {T}_{i_1\cdots i_d}\, U^\prime(\kappa)+\partial_x^2\,T_{i_1\cdots i_d}+\eta_{i_1\cdots i_d}\,.\label{eqLangevin3}
\end{equation}
Hence, $\kappa$ self averages again, but an additional contribution arises because of the Laplacian term. This equation can be formally solved using Fourier transform, with the convention:
\begin{equation}
{T}_{i_1\cdots i_d}(x,t)=:\int_{-\infty}^{+\infty} \frac{dk}{(2\pi)^{D/2}}\mathcal{T}_{i_1\cdots i_d}(k,t)\, e^{ik\cdot x}\,,
\end{equation}
where $k\in \mathbb{R}^D$ and ‘‘$u\cdot v$" is the standard scalar product. Hence, assuming again that $\Phi_c$ self averages for large $N$ around a diagonal matrix which is independent of $c$, and the dynamical equation can be easily solved after they quench as:
\begin{align}
\mathcal{T}_{i_1\cdots i_d}(k,t)=\mathcal{T}_{i_1\cdots i_d}(k,0)\rho(k,t)+\int_0^t dt^\prime\, \eta_{i_1\cdots i_d}(k,t^\prime)\, \frac{\rho(k,t)}{\rho(k,t^\prime)}\,,\label{formalsol2}
\end{align}
with:
\begin{equation}
\rho(k,t):=e^{-k^2t-g(t)}\,,\qquad g(t):=\int_0^t dt^\prime\, U^\prime(\kappa(t^\prime))\,,
\end{equation}
and the initial correlation in the Fourier space reads:
\begin{align}
\langle \bar{\mathcal{T}}_{i_1\cdots i_d}(k,0) \mathcal{T}_{j_1\cdots j_d}(k^\prime,0) \rangle_0= \Delta\, \prod_{c=1}^d \delta_{i_cj_c}e^{- \frac{k^2}{2\Lambda^2}}\delta(k-k^\prime)\,.\label{InitialFourier}
\end{align}
From \eqref{formalsol2} and \eqref{InitialFourier}, the equation for $r(t)$ becomes:
\begin{align}
r(t)=\Delta\int_{-\infty}^{+\infty} \frac{dk}{(2\pi)^{D/2}} \, e^{-2k^2(t+\tau)-2g(t)}+T\int_0^t dt^\prime\,\int_{-\infty}^{+\infty} \frac{dk}{(2\pi)^{D/2}}\, e^{-2k^2(t-t^\prime+\tau)}e^{-2(g(t)-g(t^\prime))}\,,
\end{align}
where:
\begin{equation}
\tau:=\frac{1}{4\Lambda^2}\,.
\end{equation}
We will consider firstly the quartic case because the closed equation can be solved asymptotically using Laplace transform. This in particular shows the existence of a critical temperature, which is finite in contrast to the zero-dimensional model. We then consider the general cases using a law temperature expansion, where transition temperature looks as the radius of convergence of the series.
\medskip

\subsection{The quartic case} 
Let us investigates the quartic case. If we assume \cite{Bray, LahocheSamary} that for $t$ large enough, $U^\prime(\kappa(t)) \to 0$, the self-consistent equation for $g(t)$ reads:
\begin{align}
\nonumber 0 &\approx \mu +\bar{g}_2 d\, \Bigg( \Delta\int_{-\infty}^{+\infty} \frac{dk}{(2\pi)^{D/2}} \, e^{-2k^2(t+\tau)-2g(t)}\\
&+T\int_0^t dt^\prime\,\int_{-\infty}^{+\infty} \frac{dk}{(2\pi)^{D/2}}\, e^{-2k^2(t-t^\prime+\tau)}e^{-2(g(t)-g(t^\prime))}\Bigg)\,,
\end{align}
that can be rewritten as:
\begin{equation}
G(t)= -\frac{\bar{g}_2 d}{\mu}\, \Bigg( \Delta H(t)+T F(t)\Bigg)\,,\label{closedequationtrue1}
\end{equation}
where
\begin{equation}
H(t):=\frac{1}{2^D}\frac{1}{(t+\tau)^{D/2}}\,,
\end{equation}
and:
\begin{equation}
F(t):=\int_0^t dt^\prime\, H(t-t^\prime) G(t^\prime)\,.\label{eqF}
\end{equation}
In the zero temperature limit, the problem is then solved and $G(t)\sim 1/t^{D/2}$, and then $U^\prime(\kappa(t)) \sim 1/t$, in agreement with our assumption. Note that $G(t)$ being a positive definite quantity, this makes sense only for $\mu<0$. Hence, for $t\gg \tau$:
\begin{equation}
\mathcal{T}_{i_1\cdots i_d}(k,t)=\mathcal{T}_{i_1\cdots i_d}(k,0) \left(-\frac{2^D \mu}{\bar{g}_2 d}\right)^{\frac{1}{2}} t^{D/4}\,{e^{-k^2t}} \,.
\end{equation}
The $2$-point correlation function for equal times $C(\vert x \vert,t):=\sum_I\langle \bar{T}_{I}(x,t){T}_{I}(0,t) \rangle$ behaves as $\sim e^{-\frac{\vert x \vert^2}{8t}}$ for $t$ large enough, and the correlation length grows as $\sqrt{t}$, meaning that the system does not reach a thermal regime at a finite time.
\medskip
\begin{figure}
\begin{center}
\includegraphics[scale=0.7]{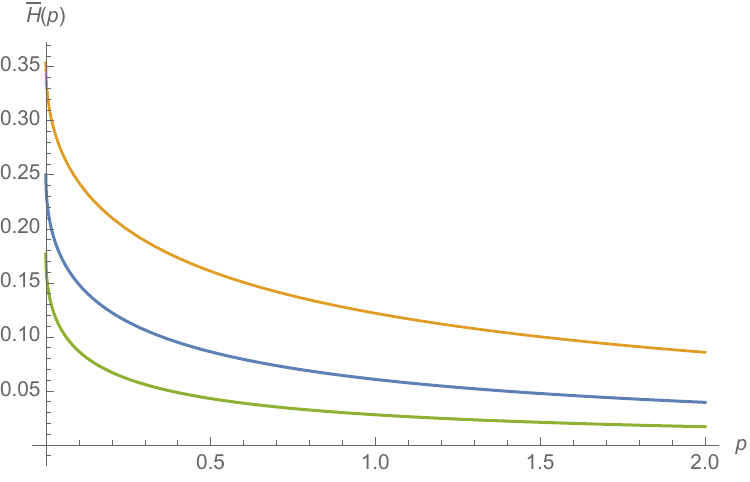}
\end{center}
\caption{Shape of $\bar{H}(p)$ for $D=3$, for $\tau=1$ (blue curve), $\tau=0.5$ (yellow curve), and $\tau=2$ (green curve).}\label{figshapeH}
\end{figure}

For $T\neq 0$, one can expect to solve the equation using Laplace transform. Let us denote as $\bar{f}(p)$ the Laplace transform of the function $f(t)$ as:
\begin{equation}
\bar{f}(p):=\int_{0}^{+\infty}dt\,f(t) e^{-pt}\,.
\end{equation}
The standard properties of Laplace transform regarding the convolution product. We get straightforwardly:
\begin{equation}
\bar{G}(p)=-\frac{\Delta}{T+\frac{\mu}{\bar{g}_2 d} \bar{H}^{-1}(p)}\,,
\end{equation}
and the Laplace transform of $H$ can be easily computed as:
\begin{equation}
\bar{H}(p)=2^{-D} p^{\frac{D}{2}-1} e^{p \tau } \int_{p\tau}^{+\infty}dt\, t^{-\frac{D}{2}}e^{-t}\,,
\end{equation}
and in particular for $D=3$:
\begin{equation}
\bar{H}(p)=\frac{2-2 \sqrt{\pi } e^{p \tau } \sqrt{p \tau }\, \text{erfc}\left(\sqrt{p \tau }\right)}{8 \sqrt{\tau }}\,.
\end{equation}
Figure \ref{figshapeH} shows the function $\bar{H}(p)$.
Hence, for large $t$ i.e. for small $p$,
\begin{equation}
\bar{H}(p)\sim\frac{1}{4 \sqrt{\tau }}-\frac{\sqrt{\pi } \sqrt{p}}{4}\,.
\end{equation}
Table \ref{table1} provides some explicit expressions for low dimensions and their asymptotic behavior for small $p$. Note that in this table, $\gamma \approx 0.57$ is the Euler constant, $\text{erfc}$ is the complement error function and $\text{Ei}(x)$ is the standard exponential integral:
\begin{equation}
\text{Ei}(x):=\int_x^\infty \frac{e^{-t}}{t}\,dt\,.
\end{equation}
At this stage, we have to clarify a technical point. The closed equation \eqref{closedequationtrue1} that we considered is only an asymptotic relation. Nerveless, we integrated it for all time taking the Laplace transform, although the closed equation is expected to be wrong for small times. Indeed, we assume the solution of the closed equation provides the true asymptotic behavior for $G(t)$, which can be motivated by the two following observations:
\begin{itemize}
\item For $t$ large enough, $H(t-t^\prime)$ suppresses low-time contributions provided that $G(t)$ has a finite limit for short times.
\item Fluctuations for $U^\prime(\kappa(t))$ are assumed to have a small standard deviation around the large time-averaged value.
\end{itemize}
We do the same assumption for higher-order potentials that we will consider in the next section. From table \ref{table1}, we show that all the computed values for $\bar{H}(0)$ are non-vanishing, but the sign of this value depends on the dimension and the size of $\tau$. For $D=1,2$, $\bar{H}(0)=\infty$, and $\bar{G}(0)$ is non singular. For $D=3$, $\bar{G}(0)$ is defined for $T<T_c$, with:
\begin{equation}
T_c(D=3)=-\frac{4\mu}{3\bar{g}_2}\sqrt{\tau} \,,
\end{equation}
which vanishes in the limit $\tau\to 0$ keeping $\mu/\bar{g}_2$ fixed. For $D>2$, we have:
\begin{equation}
T_c(D)=-\frac{\mu}{\bar{g}_2 d}2^{D-1} (D-2) \tau ^{\frac{D}{2}-1}\,.
\end{equation}

\begin{table}[htbp]
\begin{center}
\begin{tabular}{|p{1cm}||p{6cm}|p{6cm}| }
\hline
\makebox[1cm][c]{D}& \makebox[5cm][c]{$\bar{H}(p)$} & \makebox[5cm][c]{$p \ll 1$}\\
\hline \hline
\makebox[1cm][c]{1} & \makebox[5cm][c]{$\frac{\sqrt{\pi } e^{p \tau } \text{erfc}\left(\sqrt{p \tau }\right)}{2 \sqrt{p}}$} &\makebox[5cm][c]{$\sim \frac{\sqrt{\pi }}{2 \sqrt{p}}-\sqrt{\tau}$} \\\hline
\makebox[1cm][c]{2}& \makebox[5cm][c]{$-\frac{1}{4}e^{p \tau } \text{Ei}(-p \tau )$} & $\sim\frac{1}{4} (-\log (p \tau )-\gamma )-\frac{1}{4} p (\tau (\log (p)+\log (\tau )+\gamma -1))$ \\\hline
\makebox[1cm][c]{4}& \makebox[5cm][c]{$\frac{1}{16} \left(p\, e^{p \tau } \text{Ei}(-p \tau )+\frac{1}{\tau }\right)$} & $\sim\frac{1}{16 \tau }+\frac{1}{16} p (\log (p)+\log (\tau )+\gamma )$ \\\hline
\makebox[1cm][c]{5}&\makebox[5cm][c]{$\frac{4 \sqrt{\pi } e^{p \tau } (p \tau )^{3/2} \text{erfc}\left(\sqrt{p \tau }\right)-4 p \tau +2}{96 \tau ^{3/2}}$}& \makebox[5cm][c]{$\sim\frac{1}{48 \tau ^{3/2}}-\frac{p}{24 \sqrt{\tau }}$} \\\hline
\makebox[1cm][c]{6}&\makebox[5cm][c]{$-\frac{p^2 \tau ^2 e^{p \tau } \text{Ei}(-p \tau )+p \tau -1}{128 \tau ^2}$}& \makebox[5cm][c]{$\sim\frac{1}{128 \tau ^2}-\frac{p}{128 \tau }$} \\
\hline
\end{tabular}
\end{center}
\caption{Explicit expressions for $\bar{H}(p)$ in low dimension. }\label{table1}
\end{table}

The large time behavior for $G(t)$ can be deduced from the small $p$ expansion o $\bar{G}(p)$. It can be understand from standard theorems on the Laplace transform near the origin \cite{Richard}, and we have for instance the following statement:
\begin{theorem}
Let $f(t)$ be a locally integrable function on $[0,\infty)$ such that $f(t)\approx \sum_{m=0}^\infty c_m t^{r_m}$ as $t\rightarrow\infty$ where $r_m<0$. If the Mellin transformation of this function is defined and if no $r_m=-1,-2,\cdots$ then the Laplace transformation of $f(t)$ is
\beq
\bar f(p)=\sum_{m=0}^\infty c_m \Gamma(r_m+1)p^{-r_m-1}+
\sum_{n=0}^\infty Mf(n+1)\frac{(-p)^n}{n!}
\eeq
where
$Mf(z)=\int_0^\infty\, t^{z-1}f(t)dt$ is the Mellin transform of the function $f(t)$,
\begin{equation}
Mf(z):=\int_0^\infty ds \, s^{z-1} f(s)\,.
\end{equation}
\end{theorem}
For $D=3$,
\begin{equation}
\bar{G}(p) \sim \frac{\Delta}{T_c-T}+\frac{\Delta}{(T_c-T)^2} \frac{\mu}{3\bar{g}_2 } 4\sqrt{\pi } \sqrt{p} \tau+\mathcal{O}(p)\,.
\end{equation}
and for $\tau \neq 0$, $T<T_c$, we find that $G(t) \sim t^{-3/2}$. For $D=4$, the lowest order contribution behaves as $p(\log(p)+\log(\tau)+\gamma)$, and $G(t)$ behaves as $t^{-2}\log(t)$. For $D>4$, the lowest order in linear with $p$ and $G(t)\sim t^{-2}$. Hence, below the critical dimension, the relaxation time for the zero modes is infinite, and $\langle \mathcal{T}_I(0,t) \rangle \sim t^{3/4}$ for $D=3$. The $2$-point temporal correlation function for zero initial time $C(t)\sim \sum_I \,\langle \bar{{T}}_I(0,t){T}_I(0,0)\rangle $ behaves as $t^{-3/4}$, and the memory of the system follows a power law. For high temperature in contrast, the previous method breaks down, and we expect that $G(t)$ diverges faster than any power law such that Laplace transform for arbitrary small $p$ does not exist, and the system is expected to forget the initial conditions accordingly with an exponential decays.
\medskip

\subsection{Time evolution and low-temperature expansion}

\begin{figure}
\begin{center}
\includegraphics[scale=0.6]{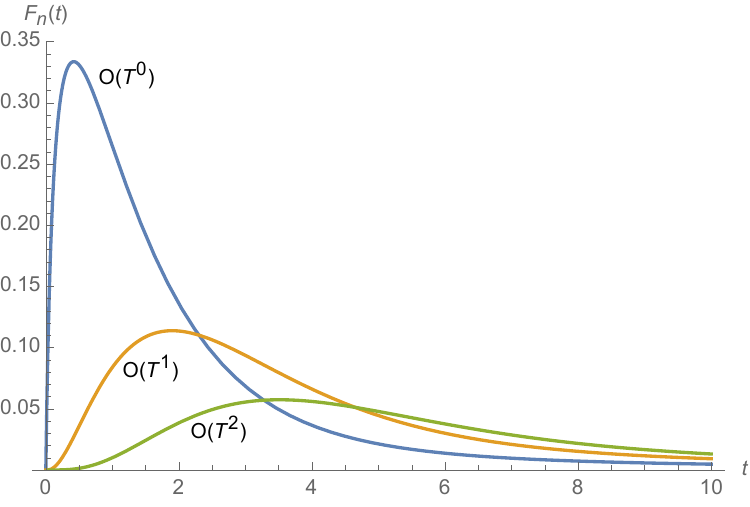}
\includegraphics[scale=0.6]{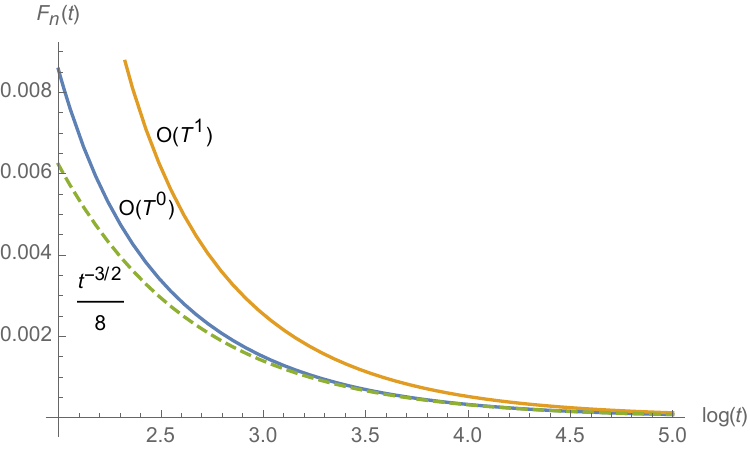}
\end{center}
\caption{On the left: Behavior of $F_0$ (blue curve), $F_1$ (yellow curve) and $F_2$ (green curve). On the right: behavior of $F_0$ and $F_1$ for late time, compared with the asymptotic behavior for $F_0(t)$ (dotted green curve).}\label{figurePlot1}
\end{figure}

\begin{figure}
\begin{center}
\includegraphics[scale=0.6]{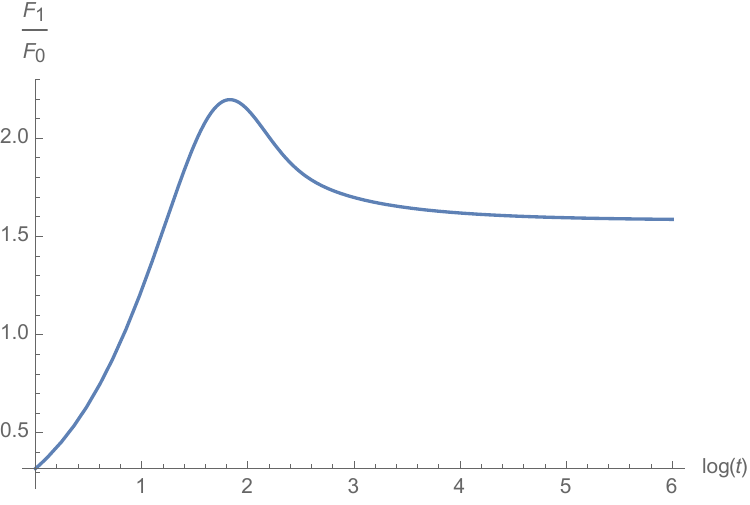}
\includegraphics[scale=0.6]{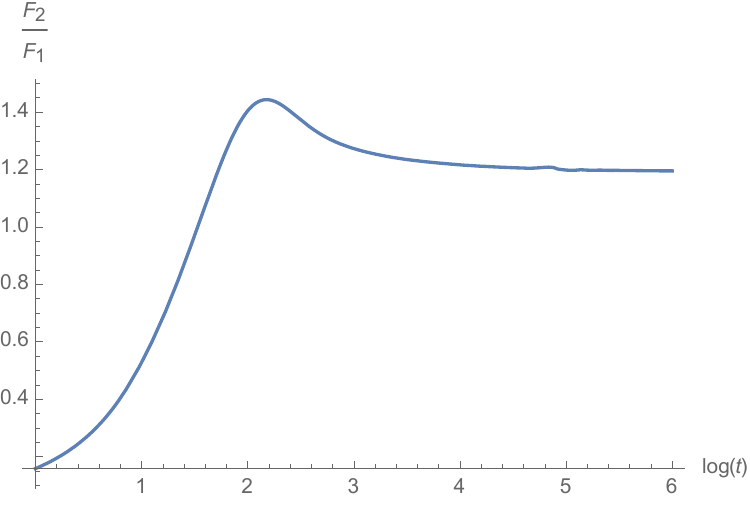}
\end{center}
\caption{Asymptotic behavior of $F_1/F_0$ (on the left) and of $F_2/F_1$ (on the right) using a logarithmic time scale.}\label{figurePlot2}
\end{figure}

As in section \ref{sec2}, we investigate the time evolution using the definition \eqref{defUprime}. In replacement of equation \eqref{equationdynamic}, we have:
\begin{equation}
\dot{G} = 2\mu G(t)+ 2d \bar{g}_2\, \left(\Delta H(t)+\, T F(t)\right)\,.\label{equationdynamic2}
\end{equation}
Taking Laplace to transform (assuming that Laplace transforms for $G$ exists), we get:
\begin{equation}
\bar{G}(p)=\bar{G}_0(p)+\bar{G}_1(p)
\end{equation}
with:
\begin{equation}
\bar{G}_0(p)=\frac{1}{p-2\mu-2d \bar{g}_2 T \bar{H}(p)}\,,
\end{equation}
and
\begin{equation}
\bar{G}_1(p)=\frac{2d \bar{g}_2 \Delta \bar{H}(p)}{p-2\mu-2d \bar{g}_2 T \bar{H}(p)}\,.
\end{equation}
assuming $\mu<0$, the series expansion in $T$ of $G_1(t)$ leads to:
\begin{equation}
\frac{G_1(t)}{\Delta}= x F_0(t)+ x^2 T F_{1}(t) + x^3 T^2 F_2(t) + \cdots
\end{equation}
where $x:= 2d \bar{g_2}$ and:
\begin{equation}
F_0(t):= \int_0^t dt^\prime\, e^{2\mu (t-t^\prime)} H(t^\prime)\,,
\end{equation}
\begin{equation}
F_{n}(t):=\int_0^tdt^\prime\, F_0(t-t^\prime) F_{n-1}(t^\prime)\,,\quad n>0\,.
\end{equation}
Figure \ref{figurePlot1} shows the behavior of $F_0$, $F_1$ and $F_2$ for $D=3$, $2\mu=1$ and $\tau=0.1$. Asymptotically, $F_0(t)$ behaves as $1/t^{3/2}$, precisely:
\begin{equation}
F_0(t)\approx \frac{0.125}{t^{3/2}}
\end{equation}
for $t$ large enough. In the second Figure in Figure \ref{figurePlot1} we show the behavior of $F_0$ and $F_1$ and the asymptotic behavior $\sim 0.125/t^{3/2}$. The functions $F_n(t)$, for, $n>0$ are also asymptotically uniformly bounded by $a/t^{3/2}$ for some $a>0$, but is can be checked numerically that $F_n(t)/t^{3/2} \to 0$ for long time. Indeed, numerically $F_{n}(t)/F_{n-1}(t) \to 0$ as $t\to \infty$, but remains almost constant for finite but late time, see Figure \ref{figurePlot2}.

\medskip

\begin{figure}
\begin{center}
\includegraphics[scale=0.6]{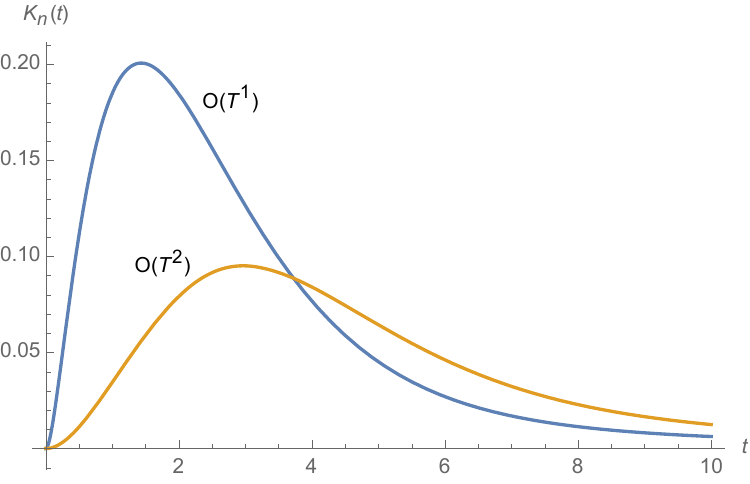}
\includegraphics[scale=0.6]{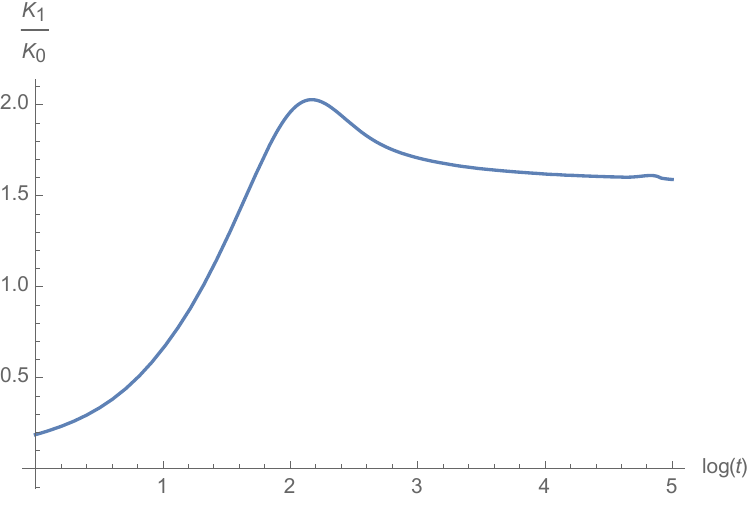}
\end{center}
\caption{On the left: behavior of $K_0$ and $K_1$. On the right: the asymptotic behavior of $K_1/K_0$ for late time using logarithmic scale.}\label{figurePlot3}
\end{figure}

For $G_0$, we get:
\begin{equation}
G_0(t)= e^{2\mu t} + x T K_{0}(t) + (xT)^2 K_1(t) + \cdots\,,
\end{equation}
where:
\begin{equation}
K_n(t):=\int_0^t dt^\prime\, e^{2\mu (t-t^\prime)} F_{n}(t^\prime)\,.
\end{equation}
The typical shape of functions $K_n$ is given in Figure \ref{figurePlot3}. Once again we find $K_0 \sim 1/t^{3/2}$ and $K_n \geq K_{n-1}$ for late time.

\subsection{Continuum limit} 

All the expressions computed above, and in particular, the transition temperatures vanish for $\Lambda \to \infty$ i.e. for $\tau \to 0$, if we keep the fractional $\mu/\bar{g}_2$ fixed. For $D=3$ for instance, this leads to the inconsistent result:
\begin{equation}
\bar{G}(p)=-\frac{\Delta}{T}\,,
\end{equation}
which has the wrong sign because $G(t)=e^{2g(t)}$ is a positive definite quantity. Indeed, one cannot take the continuum limit $\Lambda \to \infty$ without taking care of the dimensions of couplings. To make a correct dimensional analysis, it is suitable to introduce $\Omega$ such that the definitions of the model become:
\begin{equation}
\dot{T}_I(x,t)=-\Omega \frac{\delta \mathfrak{H}}{\delta \bar{T}_I(x,t)}+\eta_I(x,t)\,,
\end{equation}
where the Hamiltonian $\mathfrak{H}$ is defined as:
\begin{equation}
\mathfrak{H}[\bm T,\bar{\bm T}]:=\mathcal{H}[\bm T,\bar{\bm T}]-\sum_I\int d^Dx \, \bar{T}_I(x,t) \partial_x^2 {T}_I(x,t)\,.\label{langevineqdim}
\end{equation}
The noise distribution furthermore becomes:
\begin{equation}
\langle \eta_{i_1\cdots i_d}(x,t) \bar{\eta}_{j_1\cdots j_d}(x^\prime,t^\prime) \rangle = \Omega T\prod_{c=1}^d \frac{\delta_{i_cj_c}e^{-\frac{\vert x-x^\prime\vert^2}{2\Lambda^{-2}}}}{\Lambda^{-D} (2\pi)^{D/2}}\delta(t-t^\prime)\,.\label{distributionetabis}
\end{equation}
We introduce the bracket notation $[X]$ such that $[dx]=-[\Lambda]=-1$ and we furthermore choose the parameter $\Omega$ such that $[T]=0$. Then, because the equilibrium distribution reads $\mathrm{Exp}(-2\mathfrak{H}/T)$, we must have $[\mathfrak{H}]=0$, and we straightforwardly deduce that $[\mu]=2$ and $[\bar{g}_2]=4-D$. Finally, from \eqref{langevineqdim}, we must have $[t]=-[\Omega]-2$, and the parameter $\Omega$ is nothing but the definition of the timescale. At this step, one can repeat the analysis of the previous subsection to compute the critical temperature. Up to the replacement $H(t)\to H(t)/\Omega^{D/2}$ and $\tau \to \tau/\Omega$, it is easy to check that the expression of the critical temperature remains unchanged. Indeed, $\bar{H}(0)\sim \tau^{-\frac{D-2}{2}}$, which provides a factor $\Omega^{\frac{D-2}{2}}$ and therefore exactly compensate the factor $\Omega/\Omega^{D/2}$ arising in front of the term $T F(t)$ in equation \eqref{closedequationtrue1}. Hence, $[\mu/\bar{g}_2]=D-2$, and the equations have a relevant continuum limit only if the following scaling holds:
\begin{equation}
\frac{\mu}{\bar{g}_2 d}=\kappa_0 \Lambda^{D-2}\,,
\end{equation}
where $[\kappa_0]=0$.

\subsection{Higher order potentials} 
\begin{figure}
\begin{center}
\includegraphics[scale=0.7]{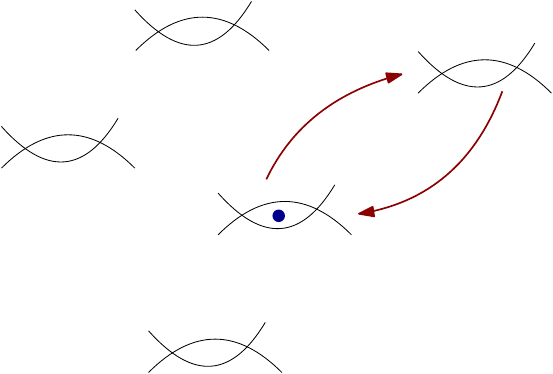}
\end{center}
\caption{The system (blue dot) frozen around local extrema for large $N$, red arrows materialize activation due to thermal fluctuations at finite $N$.}\label{vacua}
\end{figure}

The previous methods cannot be easily generalized for higher-order potentials. Indeed, for a potential of degree $m$ (hence assuming $h_m >0$ and $G(t)\neq 0$, the closed equation reads in that case:
\begin{equation}
\sum_{k=0}^m h_k (u(t))^k=0\label{equationzer0}
\end{equation}
where:
\begin{equation}
u(t):=\frac{\mathcal{A}(t)}{G(t)}\,,
\end{equation}
and:
\begin{equation}
\mathcal{A}(t,G):=
\Delta H(t)+TF(t)\,,
\end{equation}
the function $F(t)$ being defined in \eqref{eqF}. Because the coefficients $\{h_k\}$ do not depend on $t$, equation \eqref{equationzer0} has only stationary solutions, namely:
\begin{equation}
u(t)=\gamma_\mu\,,
\end{equation}
where $\gamma_\mu$ for $\mu \in \llbracket 0,m \rrbracket$ is a real and non-vanishing solution of the equation $\sum_{k=0}^m h_k \gamma^k=0$. Hence, the asymptotic solutions of equation \eqref{equationzer0} have to be such that, for large $t$:
\begin{equation}
\boxed{G_\mu(t)=\frac{\Delta}{\gamma_\mu}H(t)+\frac{T}{\gamma_\mu}\int_0^t H(t-t^\prime) G_\mu(t^\prime)\,,}
\end{equation}
where we included an index $\mu$ on $G$ refereeing to the selected zero $\gamma_\mu$. This equation looks like the equation of an effective quartic model (see equation \eqref{closedequationtrue1}), and can be solved using Laplace transform again (assuming the Laplace transform for $G(t)$ exists). We have:
\begin{equation}
\bar{G}(p)=\frac{\Delta}{\gamma_\mu \bar{H}^{-1}(p)-T}\,. \label{equationLaplaceZero}
\end{equation}
Because $G(p)$ is positive definite, this solution exists only below the critical temperature:
\begin{equation}
\boxed{T< T_c^{(\mu)}= \frac{\gamma_\mu}{\bar{H}(0)}\,. }\label{equationTc}
\end{equation}
Hence, in the large $N$ limit, each of the vacua is independent of one from the other, without overlapping, and each of them has its critical temperature looking as the radius of convergence of the low-temperature expansion. During its evolution, the system freezes around one of these vacua with a depth of order $N^{d}$, and remains around it after an expected short transition period (see Figure \ref{vacua}). Equation \eqref{equationLaplaceZero} shows that a low temperature expansion for $G(t)$:
\begin{equation}
G(t)=:\sum_{n=0}^\infty\, T^n G^{(n)}(t)\,,
\end{equation}
where $G^{(n)}(t)$ satisfies the obvious recursive relation:
\begin{equation}
G^{(n)}(p)=T\int_0^t H(t-t^\prime) G^{(n-1)}(t^\prime)\,,\quad n>0\,,
\end{equation}
does make sense only below the critical temperature, which looks like the radius of convergence of the series. Note that our derivation assumes the $G(p)$ exists for all $p$, and especially for small $p$ (i.e. for large $t$). Hence, the breakdown of the low-temperature expansion is nothing but the manifestation that this assumption for $G(t)$ becomes wrong, and then that a power time behavior leaves its place to an exponential growth i.e. the system relaxes toward equilibrium exponentially in the high-temperature regime. Finally, one expects that activation effects due to thermal fluctuations could play a role for large but finite $N$ (the famous Kramer's problem) see \cite{Kamenev,Melnikov,Berera,Chupeau,Burada,Visscher}. We investigated these aspects in a forthcoming work.
\medskip

\begin{figure}
\begin{center}
\includegraphics[scale=0.6]{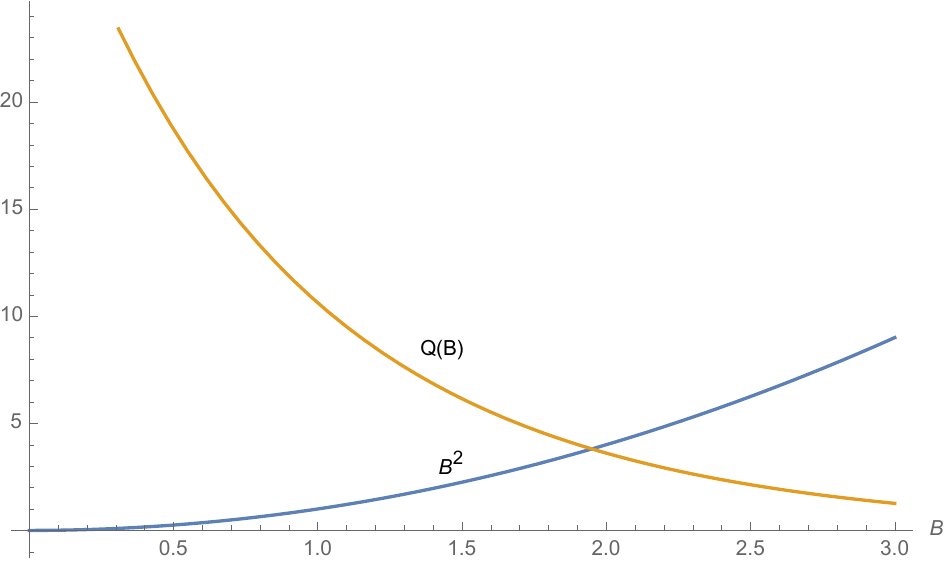}
\end{center}
\caption{Graphical solution of the equation $B^2=Q(B)$ for $\tau=0.1$ and $\gamma_\mu=0.5$ ($T$=1).}\label{figPlot9}
\end{figure}

\begin{figure}
\begin{center}
\includegraphics[scale=0.7]{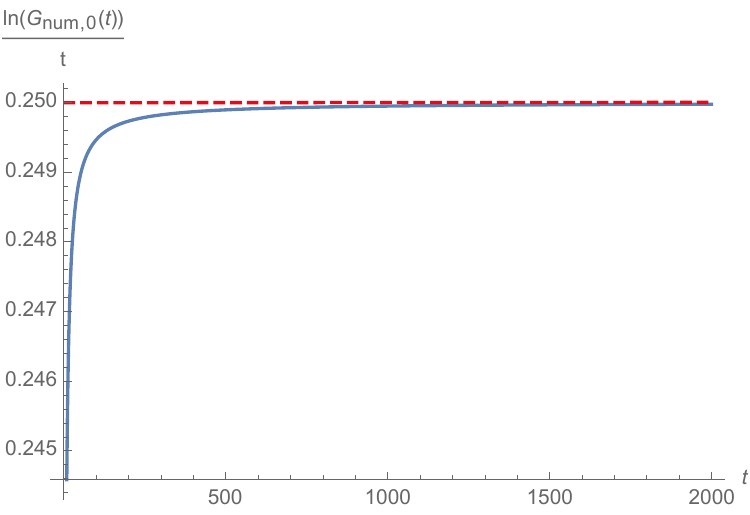}
\end{center}
\caption{Evolution of $\ln (G_{\text{num},0})/t \sim 0.25-0.054/t$ with $\Delta=0.1$, $h_m=1/2$ and $\gamma=0.5$ for $\alpha_\mu=2$.}\label{figPlot7}
\end{figure}

To conclude this subsection, let us investigate the time evolution, for the quartic model. Following \eqref{defUprime}, we have:
\begin{equation}
\sum_{k=0}^m h_k (u(t))^k=\frac{1}{2} \frac{\dot{G}}{G}(t)\,.
\end{equation}
The left-hand side can be factorized around the zeros of the polynomial, then:
\begin{equation}
h_m\prod_{k=0}^m (u(t)-\gamma_k)=\frac{1}{2} \frac{\dot{G}}{G}\,.
\end{equation}
Assuming that $u(t)$ is close to some real zero $\gamma_\mu$, one expects that $\vert u(t)-\gamma_\mu \vert \equiv \epsilon(t) \ll 1$, and expanding the left hand side of the previous equation in power of $\epsilon$, we have:
\begin{equation}
\frac{\dot{G}}{G} 
 = 2h_m R(\gamma_\mu) (u(t)-\gamma_\mu)^{\alpha_\mu}+\mathcal{O}(\epsilon^{\alpha_\mu+1}(t))
\end{equation}
where $\alpha_\mu$ denotes the multiplicity of the zero $\gamma_\mu$ and $R(\gamma_\mu)$ is the remaining contribution evaluated at $\gamma_\mu$. If we are only interested by the behavior of the system around $\gamma_\mu$, it is suitable to choose the normalization such that $R(\gamma_\mu)=1/2h_m$, and
\begin{equation}
\dot{G}(t)\approx G(t) (u(t)-\gamma_\mu)^{\alpha_\mu}\,.\label{dyngen}
\end{equation}
For $\alpha_\mu=1$, we recover exactly the analysis of the quartic case, and the system behaves as the other zeros were blinded despite thermal fluctuations. For $\alpha_\mu > 1$, the equation can be investigated numerically. Indeed, for $\alpha_\mu = 2$ and as for the $D=0$ case, $G(t)$ is an increasing function. For $T=0$, the function converges indeed exponentially as Figure \ref{figPlot7} shows, and we have again:
\begin{equation}
\frac{\ln (G_{\text{num},0}(t))}{t}=\gamma_\mu^2\left(1-\frac{\alpha}{t}\right)\,,
\end{equation}
where the index $0$ recall that $T=0$, and for $\gamma_\mu=0.5$, we get for instance $\alpha\approx 0.216$. Hence for $\alpha_\mu=2$, $G(t)$ behaves exponentially for large $t$: $G(t)\sim e^{\gamma_\mu^2 t}$, such that the equilibrium regime reduces to the equation (setting $h_m=1/2$):
\begin{equation}
(\kappa_\infty-\gamma_\mu)^2 = \gamma_\mu^2\,,
\end{equation}
and we have two solutions, $\kappa_\infty=0$ and $\kappa_\infty=2\gamma_\mu$, meaning that the systems converge towards the zero vacuum due to spatial fluctuations. This obviously contrast with the $T=0$ limit for $\alpha_\mu=1$, where $G(t)\sim 1/t^{3/2}$ and $\kappa \to \gamma_\mu$. For $T$ small enough furthermore, the low $T$ expansion for $G(t)$ shows that the exponential law holds for large $t$. Indeed, because:
\begin{equation}
\int_0^t\, dt' H(t-t^\prime) e^{a^2 t^\prime} \sim a e^{a^2 (\tau+t) } \,\Gamma \left(-\frac{1}{2},a^2 \tau \right)\,,
\end{equation}
where $\Gamma \left(n,z \right)$ is the standard incomplete gamma function:
\begin{equation}
\Gamma \left(n,z \right):= \int_z^\infty dx \, x^{n-1} e^{-x}\,,
\end{equation}
the solution $G(t)=e^{B^2 t}$ solves the asymptotic equation for $\dot{G}$ provided that $B$ satisfies the transcendental equation: $B^2=Q(B)$, with
\begin{equation}
Q(B):= \left(T B e^{B^2 \tau } \,\Gamma \left(-\frac{1}{2},B^2 \tau \right) - \gamma_\mu\right)^2\,.
\end{equation}
The equation can be solved graphically and exhibits generally a single solution, as shown in Figure \ref{figPlot9}.\\
\medskip

Note that for $\gamma_\mu$ large enough, and because for large $B$:
\begin{equation}
\Gamma \left(-\frac{1}{2},B^2 \tau \right)\approx \frac{\exp (-B^2\tau) }{(B^2\tau)^{3/2}}\,,
\end{equation}
we have: $B^2=\gamma_\mu^2(1+\mathcal{O}(T\gamma_\mu^{-3}))$. \\
\medskip

The transition temperature can be estimated from the following argument. Assuming $G(t)= \mathcal{A}(t)\, e^{\gamma_*^2 t}$ such that $\gamma_*^2=Q(\gamma_*)$, one get an equation for $\dot{\mathcal{A}}$:
\begin{equation}
\dot{\mathcal{A}}=\mathcal{A}(t)(u^2(t)-2 \gamma_\mu u(t)+\delta \gamma_\mu^2)\,,
\end{equation}
where $\delta \gamma_\mu^2:= \gamma_\mu^2-\gamma_*^2$. Hence as we just discussed, for $T$ small enough in that limit, $u(t)$ is expected to be small for late time, and we can neglect $u^2(t)$, so that the equation reads:
\begin{equation}
\dot{\mathcal{B}}\simeq-2\gamma_\mu e^{-\gamma_\mu^2 t} \left(\Delta H(t)+T \int_0^t H(t-t^\prime) \mathcal{B}(t^\prime) e^{\gamma_\mu^2 t^\prime}\right)\,,
\end{equation}
where $\mathcal{B}(t):=\mathcal{A}(t)e^{-\delta \gamma_\mu^2 t}$. The equation can be solved recursively, or by Laplace transform, which leads:
\begin{equation}
\bar{\mathcal{B}}(p)\simeq\frac{1-2\gamma_\mu \Delta \bar{H}(p+\gamma_\mu^2)}{p+2\gamma_\mu T \bar{H}(p+\gamma_\mu^2)}\,,
\end{equation}
where $\bar{\mathcal{B}}(p)$ denotes the Laplace transform of $\mathcal{B}(t)$ and we used $\mathcal{B}(t=0)=1$. For small $p$, this corresponds to an exponential decay:
\begin{equation}
\mathcal{B}(t) \sim e^{-2\gamma_\mu T \bar{H}(\gamma_\mu^2) t}\,,
\end{equation}
and the assumption $\kappa(t)\to 0$ for large $t$ requires:
\begin{equation}
T< T_c \approx \frac{\gamma_\mu}{2\bar{H}(\gamma_\mu^2)}\,.
\end{equation}

\begin{figure}
\begin{center}
\includegraphics[scale=0.6]{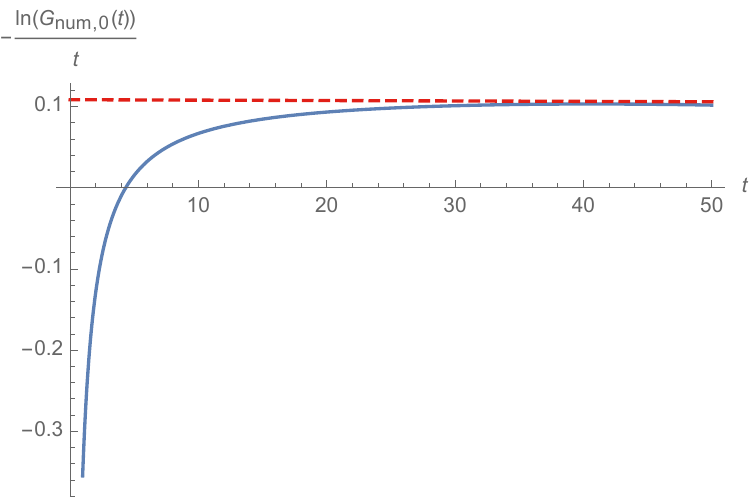}
\includegraphics[scale=0.6]{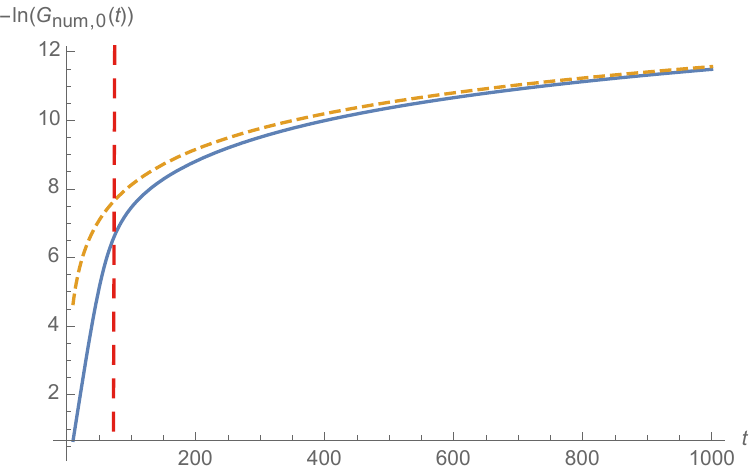}
\end{center}
\caption{Typical behavior of $G_{\text{num},0}(t)$ for $T=0$. On the left for a small time and on the right for a later time. The dotted edge on the top corresponds to the function $f(t)=1.2+(3/2) \ln (t)$.}\label{figPlot8}
\end{figure}

The same kind of investigation can be performed for other values of $\alpha_\mu$. Figure \ref{figPlot8} shows the typical behavior of $G_{\text{num},0}(t)$ (for $T=0$) for $\alpha_\mu=3$, $\gamma_\mu=0.5$ and $\Delta=1$. The evolution starts with an exponential phase, $G_{\text{num},0}(t) \sim e^{-\gamma_\mu^3 t}$ for a small time, and ends with a power law decay $G_{\text{num},0}(t) \sim t^{-3/2}$ for late time. Hence, because of the convolution property,
\begin{equation}
\int_0^t \frac{1}{(t-t^\prime+\tau_1)^{\frac{3}{2}}}\frac{1}{(t^\prime+\tau_2)^{\frac{3}{2}}} \sim \left(\frac{1}{\sqrt{\tau_1}}+\frac{1}{\sqrt{\tau_2}}\right) \frac{2}{t^{3/2}}\,,
\end{equation}
one expect that for $t$ large enough, $G_{\text{num}}(t) \sim t^{-3/2}$, as for the case $\alpha_\mu=1$. From this observation, the late time evolution can be investigated from the same method as discussed in section \ref{lowtempregime}. In particular, one can estimate the transition temperature from the following argument. During the exponential phase, the system could be close to the zero vacuum, $\kappa \sim 0$. Hence, we find for $G(t)$:
\begin{equation}
\dot{G} \approx -\gamma_\mu^3 G(t)+ 3 \gamma_\mu^2 \left(\Delta H(t)+T \int_0^t H(t-t^\prime) G(t^\prime) \right)\,,
\end{equation}
that can be solved using Laplace transform as:
\begin{equation}
\bar{G}(p)=\frac{1+3\gamma_\mu^2 \Delta \bar{H}(p)}{p+\gamma_\mu^3-T \bar{H}(p)}\,.
\end{equation}
Hence, the system ends in the non-zero vacuum $\gamma_\mu$ as $1/t$, provided that $T<T_c \approx \gamma_\mu^3/\bar{H}(0)$. Note that in this section we provided some estimations for critical temperature, but we do not prove a bound for it, an issue that should be considered in a forthcoming work.

\section{Disordered Langevin dynamics}\label{sec5}

A way to avoid the difficulty arising from UV divergences without taking care of the definition of the continuum limit is to replace the Laplacian in \eqref{eqLangevin} by a disorder coupling, materialized by a random hermitian Wigner matrix $D=D^\dagger$ of size $N^{2\times d}$:
\begin{equation}
\dot{T}_I(t)=-\frac{\partial \mathcal{H}}{\partial \bar{T}_I}-\sum_{J\in \llbracket 1,N \rrbracket^d}\,D_{IJ}T_J(t)+\eta_I(t)\,,
\end{equation}
where $T_I\equiv T_I(t)$ i.e. $D=0$ in this section. We assume that the random matrix $D$ decomposes along each colors as a sum of tensorial products:
\begin{equation}
D= \sum_{c=1}^d\, \mathbb{1}^{\otimes c-1} \otimes \sigma_c \otimes \mathbb{1}^{\otimes d-c-1}\,,
\end{equation}
where $\sigma_c$'s are hermitian Wigner matrices, which can be formally diagonalized as:
\begin{equation}
\sum_{j=1}^N\, (\sigma_{c}) _{ij}u_j^{(\lambda)}=\lambda u_i^{(\lambda)}\,,
\end{equation}
where $\{u_i^{(\lambda)}\}$ are orthogonal and normalized. We assume that $\sigma_c \in \mathrm{GUE}$ are centred gaussian matrices with the same variance $\mu$. Hence, in the large $N$ limit, the empirical distribution for eigenvalues converges toward the Wigner law:
\begin{equation}
\frac{1}{N}\sum_\lambda f(\lambda) \to \int_{-2\sigma}^{2\sigma}d\lambda\, \mu(\lambda) \, f(\lambda)\,,
\end{equation}
where:
\begin{equation}
\mu(\lambda):=\frac{\sqrt{4\sigma^2-\lambda^2}}{2\pi \sigma^2}\,.
\end{equation}
The Langevin equation can be rewritten in the eigenspace as:
\begin{equation}
\dot{T}_\Lambda(t)=-\frac{\partial \mathcal{H}}{\partial \bar{T}_\Lambda}-\left(\sum_{c=1}^d \lambda_c\right) T_\Lambda(t)+\eta_\Lambda(t)\,,
\end{equation}
where $\Lambda:=\{\lambda_1,\cdots,\lambda_d\}$ and:
\begin{equation}
{T}_\Lambda:=\sum_I T_{i_1\cdots i_d} \prod_{c=1}^d \,u_{i_c}^{(\lambda_c)}\,.
\end{equation}
In the quenched regime for $\Phi_c$, the solution for $T_\Lambda(t)$ reads:
\begin{align}
{T}_{\Lambda}(t)=&{T}_{\Lambda}(0)\rho(\Lambda,t)+\int_0^t dt^\prime\, \eta_{\Lambda}(t^\prime)\, \frac{\rho(\Lambda,t)}{\rho(\Lambda,t^\prime)}\,,\label{formalsol2}
\end{align}
where:
\begin{equation}
\rho(\Lambda,t)=e^{- g(t)-\left(\sum_{c=1}^d \lambda_c\right) t}\,.
\end{equation}
Because the spectrum is bounded, i.e. $\int d\lambda \, \rho(\lambda)=1$, no UV divergence is expected. Furthermore, because the eigenvectors $u_i^{(\lambda)}$ inherit the randomness of matrices $\sigma_c$, the condition:
\begin{equation}
T_\Lambda(0)=1\,,
\end{equation}
corresponds to a randomly distributed tensor $T_I(0)$ \cite{Leticia1,Dominicsbook}. A closed equation can be deduced from the asymptotic condition:
\begin{equation}
U^{\prime}(\kappa(t)) \to d \times 2\sigma\,,
\end{equation}
and for this section we define $U(\kappa)$ as:
\begin{equation}
U(\kappa)=(\mu+ d \times 2\sigma) \kappa + d\,\sum_{p\geq 2}\, \frac{g_p}{p}\, \kappa^p\,,
\end{equation}
such that $\mu=0$ remains the critical value. Despite these changes, equations remain essentially the same. For the quartic model, for instance, the closed equation reads again:
\begin{equation}
G(t)= -\frac{\bar{g}_2 d}{\mu}\, \Bigg( H(t)+T F(t)\Bigg)\,,\label{closedequationWig}
\end{equation}
where $G(t):=e^{2g(t)}$, $H(t)$ is now defined as:
\begin{equation}
H(t):=\int_{[-2\sigma,2\sigma]^d} \left(\prod_{c=1}^d \mu(\lambda_c)d\lambda_c\right)e^{-2\left(\sum_{c=1}^d \lambda_c\right) t-4d\sigma}\,,
\end{equation}
and $F(t)$ is still defined as
\begin{equation}
F(t):=\int_{0}^tdt^\prime\, H(t-t^\prime) G(t^\prime)\,.
\end{equation}
The closed equation is solved as well using Laplace transform and is defined for $T<T_c$, with:
\begin{equation}
T_c=-\frac{\mu}{\bar{g}_2 d} \frac{1}{\bar{H}(0)}\equiv -\frac{\mu}{\bar{g}_2 d} (2\sigma)^d\,.
\end{equation}
For higher order potential furthermore, the previous construction generalizes obviously, and we get $T_c^{(\mu)}=(2\sigma)^d\gamma_\mu$ in replacement of equation \eqref{equationTc}. As stressed above, the spectrum being bounded, no UV divergence is expected, despite the system exhibiting a non-trivial behavior for a large time. In particular, we recover that the memory of the initial condition does not vanish exponentially, and we have for the $2$-point correlation function:
\begin{equation}
C(t) \sim \frac{1}{t^{3d/4}}\,,
\end{equation}
and $U^{\prime}(\kappa(t)) \to d \times 2\sigma + \mathcal{O}(t^{-1})$ for $t$ large enough.
\\

\section{Discussions and conclusion} \label{sec7}
In this paper, we investigated the large-time behavior of a stochastic complex tensor in the cyclic melonic regime. We focused on the melonic kinetics and low-temperature regimes for different cases, including white noise limit with Laplacian or tensorial disorder, and memory effects with temporal colored noise. One of the main particularity of the melonic kinetics, which occurs for rank $d>3$ regarding kinetics for matrix or vector fields comes essentially from the ability of tensor to self-average without breaking symmetry at leading order (i.e. at the melonic order). Indeed, for vector fields, the kinetics describe a dynamical ordering resulting from the symmetry breaking for $\U(N)$ or $O(N)$ symmetry. In the melonic case, the vacuum $(\Phi_c)_{ij}=\kappa \delta_{ij}$ commute with any generator of the Lie algebra associated with this symmetry, which remains unbroken. In the case of the matrice fields, such a tractable vacuum for equilibrium dynamics is unexpected, because of the measured effect in the matrix path integral for equilibrium states, that repelled eigenvalues and leads to a non-trivial spectrum in the large $N$ limit \cite{Francesco94}. In contrast for the tensors fields, the measured effect is next to the leading order, and the eigenvalues for the intermediate fields collapse \cite{guruau2017random}. In this paper, we investigated some new aspects of the low-temperature behavior of the melonic kinetics, which is in connection with our recent contribution in the group field theory context \cite{Lahochestochastic}. Some aspects could be addressed in the future, including next to finite $N$ effects and we may implement methods like renormalization group (see  \cite{Erbin}, in preparation). Furthermore, rigorous bounds for critical temperatures should be proved, rather than estimates provided in this work.

\end{document}